\documentclass[twocolumn]{aastex62}
\graphicspath{{./}{figures/}}

\shorttitle{Planetary Nebulae major axes alignment in the Galactic Centre}
\shortauthors{S. Tan et al.}

\begin{document}

\title{When the Stars Align: A 5~$\sigma$ Concordance of Planetary Nebulae Major Axes in the Centre of our Galaxy}

\correspondingauthor{Quentin A. Parker}
\email{quentinp@hku.hk}

\author{Shuyu Tan}
\affil{The Laboratory for Space Research, Faculty of Science, \\ The University of Hong Kong, Cyberport 4, Hong Kong}

\author[0000-0002-2062-0173]{Quentin A. Parker}
\affil{The Laboratory for Space Research, Faculty of Science, \\ The University of Hong Kong, Cyberport 4, Hong Kong}

\author[0000-0002-3171-5469]{Albert A. Zijlstra}
\affil{Jodrell Bank Centre for Astrophysics, The Alan Turing Building, School of Physics and Astronomy\\ 
The University of Manchester, Oxford 
Road, M13 9PL, Manchester, UK}

\author[0000-0003-0869-4847]{Andreas Ritter}
\affil{The Laboratory for Space Research, Faculty of Science, \\ The University of Hong Kong, Cyberport 4, Hong Kong}

\author{Bryan Rees}
\affil{Jodrell Bank Centre for Astrophysics, The Alan Turing Building, School of Physics and Astronomy\\ The University of 
Manchester, Oxford Road, M13 9PL, Manchester, UK}



\begin{abstract}
\noindent
We report observations of a remarkable major axes alignment nearly parallel to the Galactic plane
of 5$\sigma$ significance for a subset of bulge 
“planetary nebulae” (PNe) that host, or are inferred to host, short period binaries. 
Nearly all are bipolar. It is solely this specific PNe population that accounts for the much weaker 
statistical alignments previously reported for the more general bulge PNe. It is 
clear evidence of a persistent, organised process acting on a measurable parameter at the heart of our 
Galaxy over perhaps cosmologically significant periods of time for this very particular PNe sample. 
Stable magnetic fields are currently the only plausible mechanism that could affect multiple 
binary star orbits as revealed by the observed major axes orientations of their eventual PNe. 
Examples are fed into the current bulge planetary nebulae population at a rate determined by their formation history and mass 
range of their binary stellar progenitors.

\end{abstract}

\keywords{binaries: close -- planetary nebulae: general -- Galaxy: bulge -- Galaxy: centre -- galaxies: magnetic fields}
\section{Introduction} 
\label{sec:intro}
The formation history of our Galactic centre, the bulge, is complex and remains under debate 
\citep[see review by][and references therein]{kormendy2004secular}. Star 
formation here occurs within collapsing molecular clouds under an environment including gravity, 
collisionless dynamics, turbulence, relativistic particle interactions and magnetic 
fields \citep[e.g.,][]{mckee2007theory}. The bulge is also permeated by magnetically confined hot 
gas \citep{blitz1993centre}. Interdependence between gravity, turbulence and magnetic fields could align a 
star's angular momentum on formation \citep[e.g.][]{crutcher2012magnetic, 2018MNRAS.473.2124G}. This may extend to binary system orbital parameters. 
For any coherence to be detectable such a process has to be strong, ordered and must
endure. Proof of this has been lacking.

However, there is a key, touchstone population that forms a highly visible component for study. These are
planetary nebulae (PNe), the short lived, ionised, gaseous ejecta from evolved low- to 
intermediate-mass stars of one to eight times the mass of our Sun \citep[e.g. see review by][]{2022PASP..134b2001K}. 
This occurs near the end of their lives as the remnant stellar core contracts, heats up and evolves 
to the white dwarf (WD) stage. The PNe mass range covers progenitor stars that may have lived for billions of years 
at the low mass end to only a few tens of millions of years at the high mass end. This is compared 
with a PN phase that lasts for only a few tens of thousands of years. 
PNe hence effectively give an instantaneous snap-shot of stellar death in a complex population like the Galactic bulge. 

PNe have key, basic shapes (morphologies) that are mostly 
elliptical, round, or bipolar, e.g. \citet{parker2006,parker2016hash}, but with some very compact 
(young and/or distant) or in rare cases, irregular or asymmetric. Only bipolars and ellipticals can 
provide a reliable orientation measurement of their principal symmetry axis from 2-D projections 
of their 3-D shapes. These can be converted into Galactic coordinates to yield Galactic position 
angles (GPAs) after determining their major axis orientations measured relative to the PN's
equatorial coordinate system. This ranges from $0^{\circ}$ to $180^{\circ}$ and is referred to as the equatorial position angle (EPA). The orientation axis from the projected bulge PN 2-D image was 
determined visually as what best represents the long symmetry axis of each PN (a subjective 
process). For elliptical PNe, the ellipse major axis was taken. For bipolar examples the geometric direction of 
the lobes and low-density interior structures were taken to determine their principle axis for the EPA measurement (see 
Fig.~\ref{fig:position_angle}). The EPAs of PNe were then converted to Galactic position angles (GPAs), 
measured from the direction of the Galactic north towards the east for further analysis following the 
formula given in \citet{corradi1998orientation} and reproduced in Section~\ref{subsec:gpa} for convenience. 
These may relate to preferred matter ejection directions during the PN phase.

\citet{Grinin1968} was the first to claim that PNe major axes may have a preferential orientation along the Galactic 
plane. This was followed by contradictory results, e.g. \citet{corradi1998orientation}. \citet{weidmann2008spatial} used a 
sample of 440 ``elongated'' PNe with measured axes and found evidence for an alignment at $\sim100^{\circ}$. 
\citet{rees2013alignment} used high-resolution \emph{Hubble Space Telescope} (\emph{HST}) and 
European Southern Observatory (ESO) New Technology Telescope imagery for 130 
compact PNe within the central $10\times10$ degrees of our Galaxy (the Galactic bulge), and found evidence at the 
3.7$\sigma$ level for alignment at a similar angle but only for a subset of bipolar PNe. Finally, 
\citet{2016IAUS..312..128D} also examined the issue and found an effect for a smaller, independently observed 
PNe sample along the broader Galactic plane whose orientation measures were based on kinematic modelling from 
3-D data cubes. The causal reasons for any alignment has not been clear: \citet{rees2013alignment} suggest strong 
magnetic fields during star formation in the Galactic bulge. 

Here, we reduce our ESO 8~m Very Large Telescope (VLT) narrow-band imagery for a slightly larger sample of 
136 compact, bulge PNe, supplemented by re-examination and re-measurement of multiple exposure, high-resolution, 
archival \emph{HST} imagery for 40 of these. This is to investigate this putative alignment 
issue afresh given the significant implications the tentative earlier results would have for our understanding 
of the pervasiveness, longevity and effect of 
ordered magnetic fields in our Galaxy. This is especially if the origin of the signal could be identified and proven real. 

The PNe sample used are considered Galactic bulge members from \citet{rees2013alignment}. 
Our new high quality imaging data yielded 
new information on large and small scale morphological features including point-symmetric structures, 
internal nebular condensations, external jets etc. These 
enabled a more detailed morphological classification following the ``ERBIAS sparm" scheme outlined in \citet{parker2006}. 
This scheme is somewhat different to the morphological classification scheme used in \citet{rees2013alignment}, 
so the bipolar samples are not quite directly comparable. Furthermore, a more accurate determination of the 
principle nebula axis of orientation was made here. Our morphological classifications, kinematic ages and detected PNe central stars for this carefully vetted sample were previously 
presented in \citet{tan2023morphologies}, hereafter Paper~I.
\section{Observations}
\label{obs}
\noindent The sample comprised 136 confirmed PNe selected to be 
highly likely physical members of the Galactic bulge within the inner 10 $\times$ 10 
degree region following rigidly applied conservative criteria set out by \citet{rees2013alignment}. 
The PNe have measured angular sizes of between 2 and 10~arcseconds. Full details are provided in Paper~I. 
The data were taken with the ESO 8.2~m VLT FORS2 instrument 
\citep{appenzeller1998successful} from programs 095.D-0270(A), 097.D-0024(A), 099.D-0163(A), and 0101.D-0192(A) 
(PI Rees and co-PI's Zijlstra and Parker). They involve short imaging exposures through narrow-band H$\alpha$ 
and/or [O~{\sc iii}] filters supplemented with H$\alpha$ images from the Wide Field and Planetary 
Camera~2 instrument (WFPC2) of the \emph{HST} of 40 objects within the sample. The \emph{HST} images were 
taken from programs HST-SNAP-8345 (PI: Sahai), HST-SNAP-9356 (PI: Zijlstra) and HST-GO-11185 (PI: Rubin).


\section{Methods}

\subsection{Selection of the PNe Samples}
For context of the full bulge PNe sample studied here (of mainly elliptical and bipolar PNe),
the binary nature of many PNe central stars has emerged as the principle shaping mechanism, especially for bipolar PNe, e.g. \citet{morris1981models, morris1987mechanisms, soker1994disks, de2009origin}. This is particularly 
for ``common envelope" situations where binary periods are shorter than a day, e.g. 
\citet{jones2017binary, boffin2019importance, ondratschek2022bipolar}, hereafter referred to as short-period binaries. 
The binary companion provides a gravitational attraction that can focus the PNe wind perpendicular to 
the binary equatorial plane and hence the observed PNe major axis position angles, e.g.
\citet{1995MNRAS.272..800H,1998ApJ...496..833S}. 

Currently, there are $\sim150$ known or suspected binary CSPN\footnote{An updated list of binary CSPNe 
is available in \url{http://www.drdjones.net/bcspn/} (accessed on 1 Mar 2023)} among the nearly 4000 
currently confirmed Galactic PNe  contained in the ``Gold Standard'' HASH database\footnote{HASH: 
online at \url{http://www.hashpn.space}. HASH federates available multi-wavelength imaging, spectroscopic and other data for all known Galactic and Magellanic Cloud PNe.}, e.g. \citet{parker2016hash,2022FrASS...9.5287P}. 
Space-based photometric surveys are used to search for short-period, close binary PNe nuclei to complement 
ground-based surveys, e.g. from the OGLE micro-lensing survey 
\citep{miszalski2009binary}, where detection is limited by atmospheric effects \citep{jacoby2021binary}. Surveys pursuing this aim were also undertaken using data from the 
\emph{TESS} \citep{aller2020planetary}, \emph{Kepler/K2} \citep{jacoby2021binary} and \emph{Gaia} 
space telescopes, e.g. \citet{gonzalez2020wide,chornay2021towards}. 

For our Galactic bulge sample of 136 PNe only 6 are in known binary systems and they are all   
short period binaries: H~2-13, H~2-29, M~2-19, M~3-42, Te~1567 and Th~3-12. One, M~1-34, 
is suspected of hosting a short period binary but doubts on correct CSPN identification
remain. Hence, it is excluded from our analysis but its GPA is consistent with our key finding 
(see later). All except one, H 2-13,  are bipolar, where bipolars constitute $\sim70\%$ of the 
full bulge sample (see Paper~I).

Furthermore, high abundance discrepancy factors (\textit{adfs}) 
$\geq$~18 have been shown by \citet{wesson2018confirmation} to be a reliable proxy for PNe hosting 
short period binary central stars. 
An \textit{adf} is the ratio of elemental abundances obtained from optical
recombination lines (ORLs) compared to abundance measurements from collisionally excited lines (CELs), e.g. \citet{corradi2015binarity, 
jones2017binary, wesson2018confirmation}. 
We carefully measured \textit{adfs} for all PNe with sufficient spectral S/N and 
measurable lines from our VLT spectroscopy. Their derivation and analysis are given Paper~IV,
Tan et al. (in prep.). We found \textit{adfs} $>$18 for 9 PNe in our sample. Of these 8 are 
bipolars and the other, H~2-42, is a round, annular PN (possibly a pole-on bipolar) that does not yield a GPA. 
We therefore add this sample to the known short-period binary PNe to provide a total sample of 14 with measurable 
GPA's for separate study compared to the full bulge PNe sample and the overall bipolars. The GPAs for PNe with known and 
inferred short-period binaries are presented in Table~\ref{bgpas}.
\setlength{\tabcolsep}{2.5pt}
\renewcommand*{\arraystretch}{1}
\begin{table*} 
\centering
\caption{Binary periods (in fractions of a day) or \textit{adf} values, GPAs (as defined in Section~\ref{subsec:gpa}),
morphologies (following the``ERBIAS sparm'' classification scheme described in \citet{parker2006}  and illustrated in Fig.~3 of Paper~I, where ''B" means the PN is bipolar and ``E" elliptical with lower case letters being sub-classifiers), positions, distances (when available in kpc), outer radii and kinematic 
ages (in kiloyears) of the short-period 
(or inferred short-period) binary PNe sample. Part \textbf{(a)} presents the 6 PNe with 
observationally confirmed short-period binary central stars plus M~1-34$^{\bullet}$ (strongly suspected of 
being a short-period binary system but excluded from the analysis). Part \textbf{(b)} lists the 8 PNe exhibiting a 
high \textit{adf} ($\geq$~18), as determined from Paper~IV, that have a measurable GPA to go into the analysis.}
\label{bgpas}
\begin{tabular}{lccccccccc}
\multicolumn{10}{c}{\textbf{(a)}}\vspace{0.1cm}\\
\hline
\multirow{2}*{PN G} & \multirow{2}*{Name} & Period &    GPA & \multirow{2}*{Morph.}&  
\multirow{2}*{RAJ2000}  &   \multirow{2}*{DECJ2000} &   D &  $R\mathrm{_{out}}$  &                 
$t\mathrm{_{kin}}$ \\ & & [days] & [$^{\circ}$] & & & &[kpc] & [pc] & [kyr]\\
\hline
000.2$-$01.9 &      M~2-19 &\phantom{ai} 0.670$^{1}$ \phantom{ia}&  165.0 &Brs &  17:53:45.64 &  -29:43:47.0 &  
6.49$_{-1.16}^{+1.27}$&   0.31 &   21.6$_{-4.4}^{+4.7}$ \\
002.8+01.8 &  Te~1567 & 0.171$^{1}$ &  108.0&Bars &  17:45:28.34 &  -25:38:11.9 &         
$-$ &   0.18 &    $\ $6.2$_{-1.9}^{+4.4}$ \\
356.8+03.3 &      Th~3-12 &0.264$^{1}$ &   94.5&Bp &  17:25:06.12 &  -29:45:17.0 &      
$-$ &   0.07 &      $\ $5.4$_{-0.7}^{+0.8}$ \\
357.2+02.0 &       H~2-13 & 0.897$^{1}$ &   91.0&Emrs &    17:31:08.11 &   -30:10:28.2 &
$-$ &   0.09 &     $\ $7.8$_{-0.8}^{+1.7}$ \\
357.5+03.2 &       M~3-42 & 0.320$^{1}$&   92.5&Bas &    17:26:59.85 &   -29:15:32.7 &  
$-$ &   0.20 &                  $\ $5.1$\pm$0.7 \\
357.6$-$03.3 &       H~2-29 &0.244$^{2}$ &   94.0& Bs &    17:53:16.82 &   
-32:40:38.5 &  5.37$_{-2.92}^{+2.99}$ &   0.16 &    $\ $7.4$_{-3.9}^{+4.3}$ \\
357.9$-$05.1 &       $^{\ }$M~1-34$^{\bullet}$ & $\ \ -\ \ ^{2}$ & 92.5 & Bmps &    
18:01:22.20 &   -33:17:43.1 &     $-$ &   0.51 &  17.8$_{-5.4}^{+11.7}$ \\
\hline  \vspace{-0.14cm}
\end{tabular}

\begin{tabular}{lccccccccc}
\multicolumn{10}{c}{\textbf{(b)}} \vspace{0.1cm}\\
\hline \multirow{2}*{PN G} & \multirow{2}*{Name} &  \multirow{2}*{\textit{adf}} &   GPA  & \multirow{2}*{Morph.}  &  
\multirow{2}*{RAJ2000}  &   \multirow{2}*{DECJ2000} &   D &  $R\mathrm{_{out}}$  &         
$t\mathrm{_{kin}}$ \\ & &  & [$^{\circ}$] & & & &[kpc] & [pc] & [kyr]\\
\hline
000.2$-$04.6 & Sa~3-117 & 19.0$\pm1.2$ & 114.5 &Bs & 18:04:44.09 &  -31:02:48.9 &                 $-$ &   0.15 
&   5.2$_{-1.6}^{+3.5}$ \\
001.3$-$01.2 &     Bl~M & 178.4$\pm52.4$ &  102.0 &Bmrs & 17:53:47.16 &  -28:27:17.8 &  
4.84$_{-2.84}^{+3.77}$ &   0.07 &   2.4$_{-1.6}^{+2.9}$ \\
002.7$-$04.8 &  M~1-42 & 18.5$\pm1.4$  &   97.0 &Bamprs &   18:11:04.99 &   -28:58:59.2 &   
4.31$_{-1.03}^{+1.10}$ &   0.19 &   6.9$_{-1.7}^{+1.8}$ \\
005.0$-$03.9 &       H~2-42 & 147.1$^{+20.5}_{-23.2}$ &   $-$ & Ramrs &  18:12:22.99 &  -26:32:54.5 &  $-$ &   
0.25 &   8.7$_{-2.6}^{+5.9}$ \\
007.6+06.9 &       M~1-23 & 19.0$\pm1.5$ &  111.5 & Bamrs &  17:37:22.00 &  -18:46:42.0 &                 $-$ &
0.32 &  11.1$_{-3.4}^{+7.2}$ \\
353.2$-$05.2 &       H~1-38 &42.0$\pm2.7$ &  102.5&Bmrs &    17:50:45.20 &   -37:23:53.1 &                 $-$ 
&   0.33 &  11.5$_{-3.4}^{+7.6}$ \\
357.1+04.4 & Terz~N~18 & 19.3$^{+4.3}_{-3.2}$ & 127.5 & Ers & 17:21:37.98 & -28:55:14.6 & $-$ & 0.21 & 
7.3$^{+4.7}_{-2.2}$ \\ 
357.9$-$03.8 &       H~2-30 & 184.2$^{+26.1}_{-25.8}$ &  117.5 &Bmrs &    17:56:13.93 &   -32:37:22.2 &  
8.31$_{-1.66}^{+1.75}$ &   0.26 &   9.0$_{-3.0}^{+6.4}$ \\
359.6$-$04.8 &       H~2-36 & 51.9$^{+3.5}_{-3.6}$ &  117.0 &Brs &  18:04:07.75 &  -31:39:10.7 &               
$-$ &   0.32 &  11.1$_{-3.3}^{+7.2}$ \\
\hline
\end{tabular}
\\
\tablenotetext{}{References. $^{1}${\cite{jacoby2021binary}}; $^{2}$\cite{miszalski2009binary}.}
\end{table*}

\subsection{Measurement of PN position angles}
\label{subsec:gpa}
\begin{figure}
\centering
\includegraphics[width=0.22\textwidth]{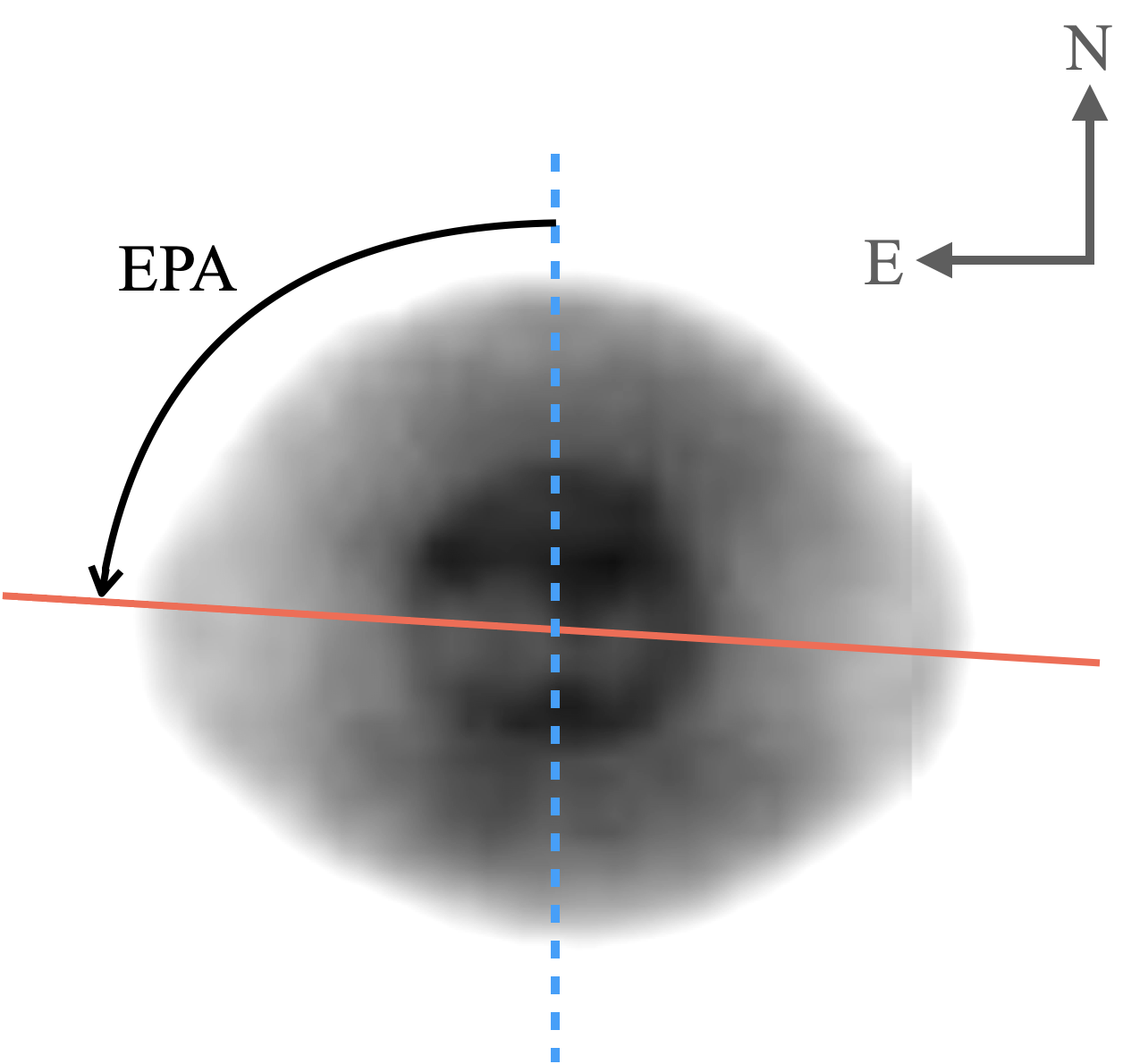}
\hspace{0.8cm}
\includegraphics[width=0.185\textwidth]{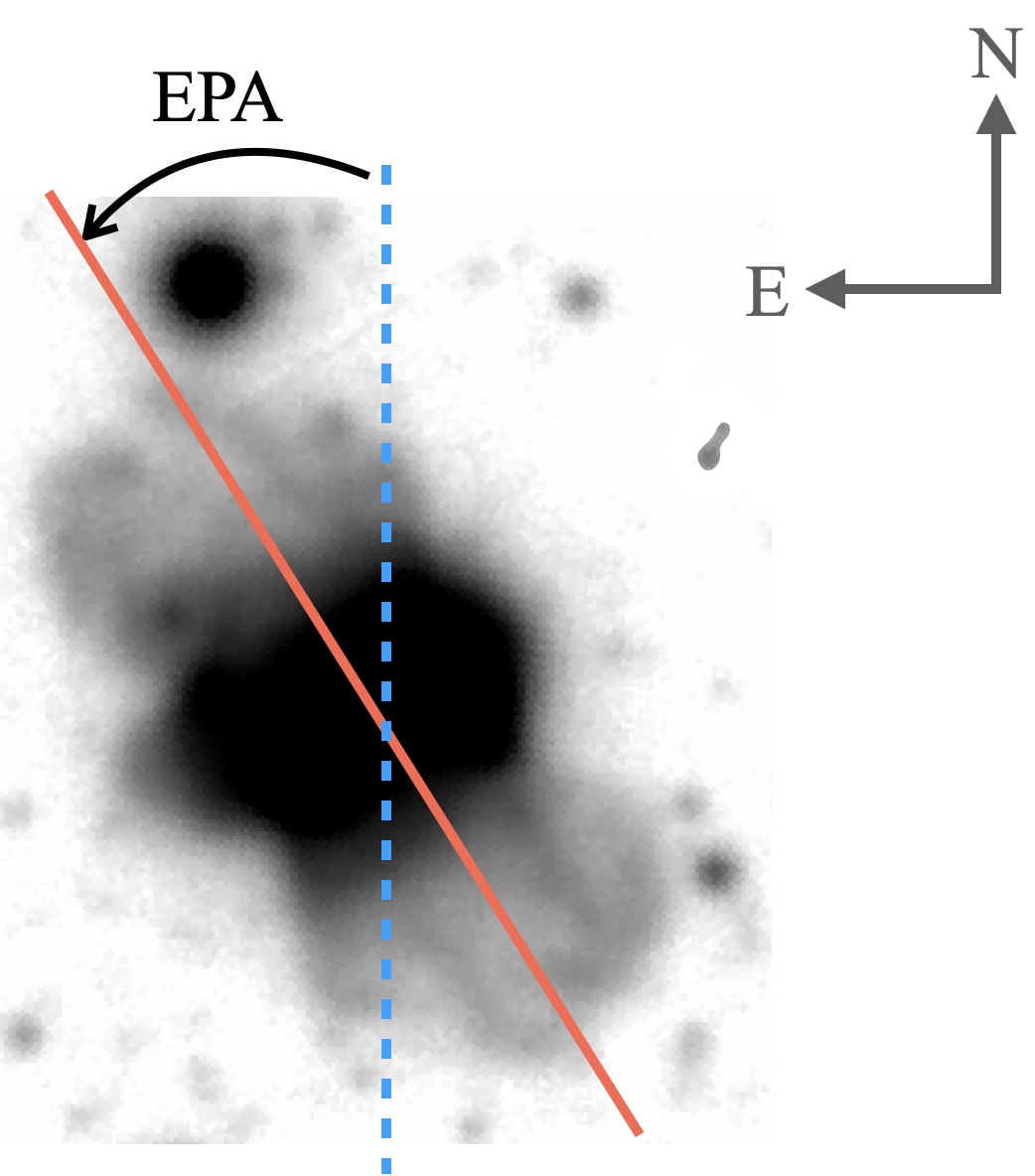}
\caption{Measurement of the equatorial position angle (EPA). The orientation axis measured from the projected 2-D image of 
a bulge PN was determined visually and the axis that best represents the long symmetry of each PN. The left panel shows a 
typical elliptical PN M~1-20 with a morphology of ``Ems" and an EPA of $87^{\circ}$ according to its 
major axis from \emph{HST} observations. The right panel shows bipolar M~1-34 from our sample with a 
``Bps" morphology and an EPA of $31^{\circ}$ measured from the VLT observation.}
\label{fig:position_angle}
\end{figure}

\noindent From the combined VLT and \emph{HST} data of the 136 PNe, a total of 126 could provide a meaningful
position angle with the other ten being too round and/or compact. 
This angle is as measured between the major axis of the apparent nebular 
elongation from the north towards the east, as shown in 
Fig.~\ref{fig:position_angle}. Orientations are measured relative to the PN's
equatorial coordinate system from $0^{\circ}$ to $180^{\circ}$, 
also referred to as the equatorial position angle (EPA). The orientation axis measured from 
the projected 2-D PN image was determined visually and the axis that best 
represents the long symmetry of each PN was estimated, taking into account low-density interior 
structures. EPAs were rounded to the nearest $0.5^{\circ}$. To determine uncertainties, 50 general Bulge PNe 
were measured independently by three authors (ST, AR \& QAP). This gave a typical difference of $\sim2$-$5^{\circ}$ 
and is considered the random error in our EPA measurements. 

The EPAs were converted to Galactic position angles (GPAs) using our measured geometric centroid 
positions in Galactic coordinates for further analysis. This is following the formula 
in \citet{corradi1998orientation} and reproduced below for convenience. 
This transformation is done by computing the angle $\psi$ subtended at each object's position by 
directions of both the equatorial and Galactic north. Using standard relations for spherical triangles, 
for epoch $\mathrm{J} 2000.0$ $$\psi=\arctan \left[\frac{\cos \left(l-32.9^{\circ}\right)}{\cos b \cot 62.9^{\circ}-\sin b 
\sin \left(l-32.9^{\circ}\right)}\right],$$ where $l$ and $b$ are the Galactic coordinates of the object. 
The corrections of $\psi$ for precession of the equinoxes are negligible in the present context. 
The  GPA is then defined as the nebular axis position angle 
measured from the direction of the Galactic north towards the east, 
$$ \text{GPA}=\text{EPA}+\psi.$$ The uncertainty in GPAs takes the same value as for the EPAs. 
EPA measurements for the full bulge PNe sample and indeed for confirmed PNe across the rest of the 
Galaxy (when measurable) will soon be available in Ritter et al.~(in prep).

\subsection{Statistical Techniques and Analysis}
Circular data, as concerned in this study, is a form of directional data in Statistics and a range of texts have been 
dedicated to describing their special statistical treatment: e.g. \citet{fisher1995statistical, mardia2000directional, 
jammalamadaka2001topics, pewsey2013circular}, and more recently, \citet{ley2017modern, pewsey2018applied}. To investigate 
any GPA alignment signal in our carefully remeasured bulge PNe samples and identify the responsible group, if any, we employed a 
range of established frequentist tests of circular uniformity on both the entire sample and different subgroups. 
These subgroups included bipolar PNe, bipolar PNe without confirmed or suspected short-period binaries, 
all PNe regardless of morphology but excluding the short-period binaries and just the short-period binary PNe. 
Additionally, we performed Monte Carlo simulations for drawing random samples to test the null hypothesis of 
uniform distributions by estimating the likelihood of obtaining the observed GPA spread for different subgroups. 
Further, we used Bayes factors to quantify the evidence in the data in support of a null hypothesis of circular 
uniformity versus an alternative hypothesis of a von Mises distribution \citep[i.e. circular analogue of the 
normal distribution,][]{vonMises1918}. Finally, we used a separate Markov Chain Monte Carlo (MCMC) analysis to 
fit a von Mises distribution to the observed data when the Bayes factor indicated strong evidence for a von Mises
distribution over a uniform distribution.
\subsubsection{Frequentist methods} 
A common statistical exploration of circular data involves testing for the presence of unimodal bias in the distribution 
around the circle (i.e., a concentration of data in a specific region) or determining if the null hypothesis of a 
uniform spread throughout the circle is supported by the underlying population. A range of statistical tests have 
been proposed for this purpose \citep[see][for a review]{batschelet1981circular}, and applied in previous studies 
to assess potential biases in the distribution of GPAs in PNe. In our analysis, we used a Rayleigh test, a Kuiper 
test, a projected Anderson-Darling (PAD) test and a Watson test for a null hypothesis of circular uniformity 
against an alternative hypothesis of non-uniformity on the doubled GPAs. The statistics were primarily implemented 
in \texttt{R} \citep{R_language} with \texttt{unif\uline{ }test} of the \texttt{sphunif} package \citep{sphunif} 
and \texttt{r.test} of the \texttt{CircStats} package \citep{berens2009circstat} for the Rayleigh test. 
The $p$-values were estimated using the exact null distributions, which were approximated through Monte Carlo simulations. The Rayleigh test is based on the length of the mean vector and is 
consistent against uni-modal alternatives, e.g. \citet{rayleigh1919xxxi, fisher1995statistical, mardia2000directional}. The Kuiper and Watson's 
tests are based on the maximum and mean differences 
between empirical and hypothesized uniform cumulative distribution functions \citep{kuiper1960tests, batschelet1981circular}. They 
are consistent against all alternatives to uniformity and more sensitive to departures from uniform and multi-modal distributions 
than the Rayleigh test \citep{batschelet1972recent}. The robustness of the projected Anderson-Darling test used was proven with a 
simulation study by \citep{garcia2023projection} and is a good reference test as it has excellent performance against uni-modal and non uni-modal distributions.

\subsubsection{Monte Carlo simulations and Bayesian analysis}
To address potential inaccuracies in approximating $p$-values in statistical tests, which may occur with small sample sizes \citep{ajne1968simple, freedman1981watson, arnold2011nonparametric}, 
we employed Monte Carlo simulations as an supplementary 
approach alongside frequentist tests to estimate the likelihood ($p(H_{\mathrm{0}}|\theta)$) of obtaining the observed GPAs from a uniform 
circular distribution. We randomly selected the same number of angles as in the data from a uniform distribution between $0^{\circ}$ and 
$180^{\circ}$ $10^{8}$ times and then calculated the fraction that produced a standard deviation ($\sigma_{\mathrm{{GPA}}}$) equal to or 
smaller than that of the observed data. On this basis, our short-period binary 
subsample shows an extremely low likelihood of a uniform distribution of GPAs (see results section below). 

To further investigate the presence of a concentration of GPAs of our samples, we applied a Bayesian hypothesis test 
\citep{mulder2021bayesian} for circular uniformity against a von Mises alternative based on the Bayes factor, the ratio of two marginal 
likelihoods, on the doubled GPAs. This was computed using the \texttt{circbayes} package 
\citep{circbayes}. Flat priors for the mean direction, $\theta_{\mathrm{\mu}}$ and the concentration parameter, $\kappa$ 
(1/$\kappa^{1/2}$ is equivalent to the $\sigma$ of a normal distribution) were used. The Bayes factors are presented on a log scale in 
column~7 of Table~\ref{tb:gpa_stats}. We follow the standard ``rules of thumb'' for evidence thresholds 
outlined in \citet{kass1995bayes}. Here, Bayes factors between 20 and 150 ($1.3 < \log_{\mathrm{10}}B < 2.2$)  are considered 
``strong", while Bayes factors greater than 150 ($\log_{\mathrm{10}}B > 2.2$) are labelled as ``very strong" evidence for the 
alternative hypothesis against $H_{\mathrm{0}}$.  

We then applied the MCMC approach to fit a mixture of von Mises distribution to GPAs of the short-period binary sample with 
\texttt{emcee} \citep{foreman2013emcee,foreman2019emcee}, an ``affine" invariant MCMC implemented in \texttt{Python} 
\citep{van1995python}. The chosen von Mises distribution has two antipodal modes in 0$^{\circ}$ and 180$^{\circ}$, 
respectively, to correspond to the practice of our GPA measurement using flat priors where the prior of $\kappa$ 
truncates at 30. We used 500 parallel samplers and 10,000 steps per sampler, with a burn-in period of 6500 steps in all cases. The step 
size was chosen to adequately sample all parameter spaces based on an analysis of the auto-correlation function of our data as discussed 
in the \texttt{emcee} documentation\footnote{\texttt{emcee} documentation on autocorrelation analysis and convergence can be found at: 
\url{https://emcee.readthedocs.io/en/latest/tutorials/autocorr}}.

\section{Summary Results}
We present statistical analysis results for the five subsets from our bulge PNe sample and summarized in 
Table~\ref{tb:gpa_stats}. The first subset includes all 126 PNe in our sample with a measurable GPA. The second subset 
consists of 112 PNe that do not host short-period binaries, irrespective of their morphology. The third subset contains all PNe 
classified as bipolar (93 PNe), while the fourth subset include 81 bipolar PNe but excluding those that also host or are 
inferred to host short-period binaries. Finally, the fifth subset consists only of 14 PNe hosting short-period binary central 
stars \cite[i.e. they have measured short-period binary CSPN or have \textit{adfs} derived from our high S/N, VLT PNe emission line 
spectroscopy that have been shown are a strong proxy for PNe hosting short period binaries,][]{wesson2018confirmation}. 

\begin{figure}
    \centering
   \includegraphics[width=0.41\textwidth]{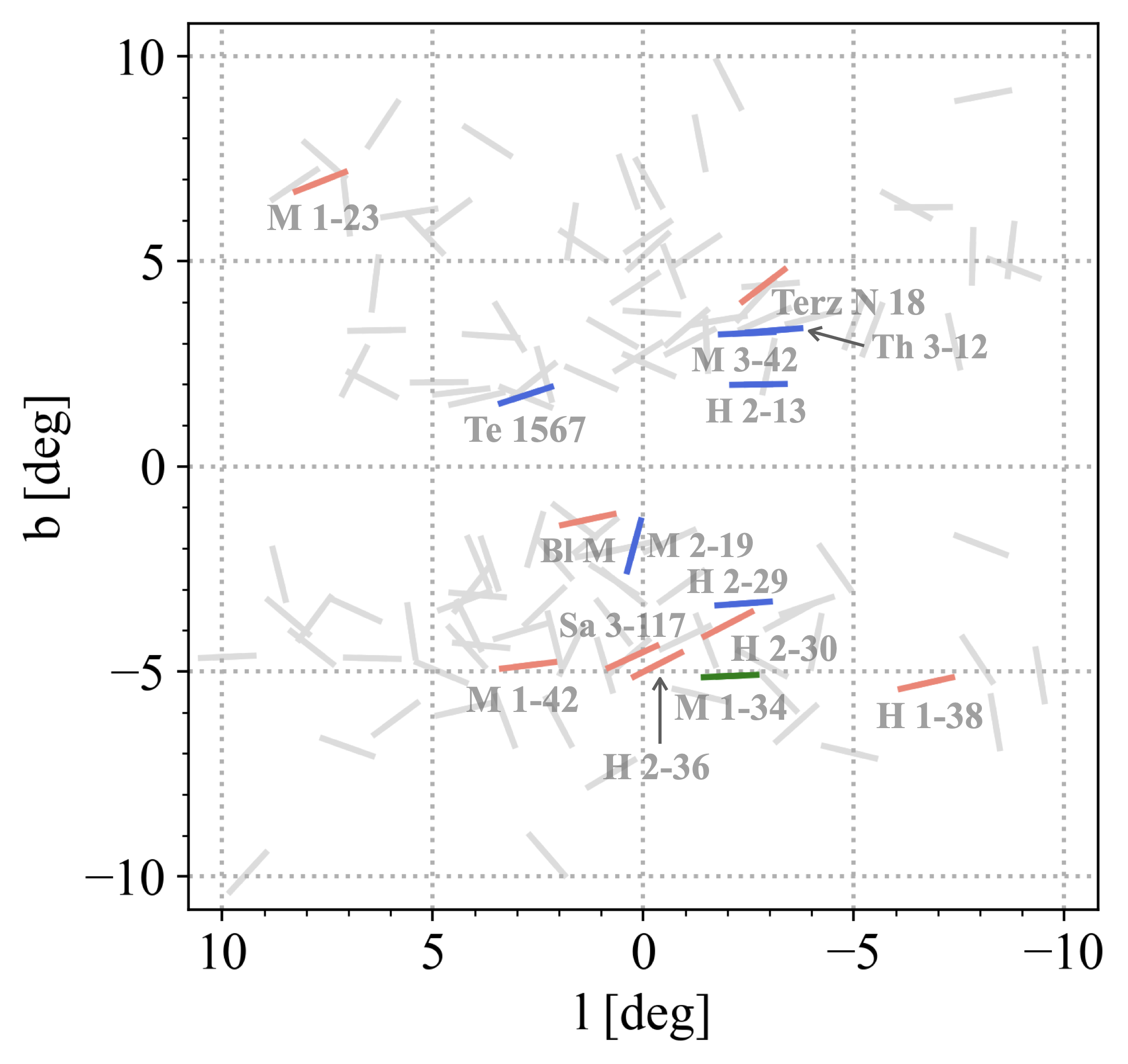} 
    \caption{The bulge distribution of measured GPAs for all PNe in our sample. The grey symbols are for 
    the general sample while those PNe with confirmed binary central stars are blue vectors and
    those with high \textit{adfs} (that are taken as a reliable proxy for PNe hosting
    short-period binaries) are pink vectors. M~1-34, suspected of hosting a short period binary, is plotted with
    a green vector. The usual name of each PNe are plotted under each blue, pink or green vector.}
    \label{fig:GPA_dist}
\end{figure}

Figure \ref{fig:GPA_dist} displays the Galactic coordinate distribution for all bulge PNe in our sample with reliable GPA 
measurements, as well as for the subset of PNe hosting or likely to host short-period binary central stars. The plotted vector 
orientations indicate the measured GPA, with the six PNe with confirmed short-period binary CSPNe and the eight PNe with high 
\textit{adfs} $\geq$~18 plotted as blue and pink vectors, respectively. The non-binary bulge PN sample is represented by grey 
vectors. The PN M~1-34, suspected of hosting a short-period binary, is plotted with a green vector. Excluding M~1-34, the combined 
assumed short-period binary sample has a mean GPA of 107.5$^{\circ}$, with a narrow $\sigma=17.5^{\circ}$, indicating a direction 
nearly parallel to the Galactic plane. In contrast, the non-binary sample has a mean GPA of $82.0^{\circ}$, with a large 
$\sigma=60.0^{\circ}$ for n~=~112.  Images of all these short-period binary PNe overlaid with the measured GPA vectors are 
provided in Appendix~A.

In Fig.~\ref{fig:stats_plot}, we contrast graphical representations for the 126 full PNe sample and the 14 PNe binary sub-sample for 
which GPA measures were possible. These include finger plots of GPA distribution, rose plots (a circular form of bar chart) of 
the doubled angles and the quantile–quantile (Q–Q) plots showing deviation from uniformity. The strong, non-uniformity of the GPA's of 
the short-period binary sample is obvious. We conducted a Monte Carlo simulation (see Methods) where the estimated likelihood to 
obtain a $\sigma\leq$ 17.5$^{\circ}$ as observed is only 0.0005\%. We also applied a Rayleigh test, a Kuiper test, a projected 
Anderson-Darling test and a Watson test for a null hypothesis of GPA circular uniformity (refer to Methods). The $p$-values, or 
probabilities that the null hypothesis is true, are listed in columns~2-5 in Table~\ref{tb:gpa_stats} for all four statistical tests. 

\noindent


The results also clearly indicate that the general PNe samples of non short-period binary PNe (regardless of 
morphology) are highly likely to be uniformly distributed.  Furthermore, the bipolar sample of 81 PNe that excludes 
the short period binary sub-sample show no significant departures from random GPAs. 

Our specific short-period binary sample however has a highly significant Rayleigh test statistic
and its rose plot in Fig.~\ref{fig:stats_plot}\textbf{(f)} show no multi-modal features that could lead to a type~II error 
(i.e. failure of rejection of the null hypothesis when it is actually false). The Rayleigh test result is reliable, 
showing a high concentration around a preferred angle direction. The results reported for the the Kuiper and Watson's tests 
are also extremely significant in terms of $\sim5~\sigma$ departures from normality for the short-period binary sample 
which are also almost exclusively bipolar. The weak alignment signal previously reported in 
\citet{weidmann2008spatial} and even the stronger alignment significance reported 
by \citet{rees2013alignment} for their bipolar sample were both diluted. The actual signal in our study 
arises solely from a very specific but relatively small PNe sub-sample of short period binary PNe. 
This is our major discovery. 
 
We also applied a Bayesian hypothesis test using non-informative flat priors (see Methods) to each sample. 
The Bayes factors for the general bulge PNe sample and the other samples (but excluding the short-period binary 
sub-sample) provide positive evidence for the alternative hypothesis of a von Mises unimodal circular distribution. 
For our full bipolar subsample the Bayes factor does give some weak support for a von Mises distribution c.f. a uniform 
distribution with $\log_{\mathrm{10}}B =1.23$. However, the Bayes factor for the short-period binary subsample is a remarkable 
$\log_{\mathrm{10}}B =5.5$ and provides extremely strong preference for a concentrated von Mises distribution. 

Through an MCMC analysis, we find the most likely values for the orientation and standard deviation of the binary 
sub-sample are $109.0\pm 4.9$ and $15.0^{+6.5}_{-2.8}$ degrees, respectively. The presented lower and upper errors 
correspond to the $16^{\mathrm{th}}$ and $84^{\mathrm{th}}$ percentiles of the samples, which coincides with 1~$\sigma$ 
confidence levels in the case of a Gaussian posterior distribution. The associated ``corner plot'' is 
presented in Fig.~\ref{fig:corner_plot} of Appendix~B, displaying the results of MCMC parameter estimation of the von Mises 
model applied to our short-period binary PN sample. Additionally, we provide a comparison between 
the model estimations and the observed data in Fig.~\ref{fig:mcmc_fits}. The alignment of GPAs of our 
short-period binary subsample is further supported by the observed high concentration parameter 
$\kappa$. 
 
Hence, regardless of the statistical approach applied, the GPAs of our short-period binary PNe sub-sample are highly concentrated.
This is the first time preferred PNe major axes alignments have been demonstrated with such high statistical power but only for a special 
sub-sample of PNe that host, or are inferred to host, short-period binary CSPN. 
These are almost exclusively bipolars ($\sim80\%$) and so are likely also to be undergoing 
common envelope evolution (CEE), e.g.  \citet{ivanova2013common}. What could be the explanation? 

One key consideration is at what evolutionary stage is the effect imposed? 
If the effect reflects today a respect for a binary
orbit inclination ``frozen in" at the time of the original binary system formation then an enduring longevity of 
the causality must exist. This is because such systems did not all form at the same time but over hundreds of 
millions of years and perhaps billions of years ago. If the effect is imposed very recently on all these short lived 
PNe visible now, and in just those systems where the binary orbits have decayed to short periods 
and on entering a CEE phase, then what fast acting mechanism can do this?

\begin{table*}
\centering
\caption{Statistical analysis summary results. Listed are the $p$-values given by various statistical tests, likelihood 
$p({H_{\mathrm{0}}}|{\theta})$ values estimated from Monte-Carlo simulations and Bayes factors, B (in $\log_{\mathrm{10}}$) for a von Mises circular distribution against a uniform distribution for the doubled 
GPAs from the full PNe sample, the non short-period (SP) sample, the bipolar sample, the non SP bipolar sample 
and the SP binary PNe. The equivalent Gaussian "sigma" levels for $p$-value significance and likelihood values 
are shown in brackets. The $\sim5\sigma$ significance results for the SP sample is remarkable.}
\label{tb:gpa_stats}
\begin{tabular}{p{3.8cm}rrrrrc}
\hline
Sample& $p(\mathrm{Rayleigh})$ & $p(\mathrm{Kuiper})$  &  $p(\mathrm{PAD})$ & $p(\mathrm{Watson})$ & 
$p({H_{\mathrm{0}}}|{\theta})$ & $\log_{\mathrm{10}}$B \\ 
\hline  All (126)& 0.024 [2.3] & 0.0062 [2.7]  & 0.13 [1.5] & 0.013 [2.5] & 0.024 [2.3] & 1.1 \\ 
 Non-SP Binary (112)& 0.26 [1.1] & 0.13 [1.5] & 0.99 [0.0] & 0.22 [1.2] & 0.26 [1.1] & 0.26 \\ 
B (93) & 0.023 [2.3] & 0.022 [2.3] & 0.085 [1.7] & 0.014 [2.5] & 0.022 [2.3] & 1.2 \\
B, Non-SP Binary (81) & 0.17 [1.4] & 0.17 [1.4] & 0.84 [0.2] & 0.18 [1.3] & 0.17 [1.4] & 0.57 \\
SP Binaries (14)& 5.0$\times10^{-6}$ [4.6] & 1.2$\times10^{-6}$ [4.9] & 2.8$\times10^{-6}$ [4.7]  & 2.9$\times10^{-6}$ [4.7] &  
5.0$\times10^{-6}$ [4.6] & 5.5 \\ \hline 
\end{tabular}
\end{table*}
\begin{figure*}
    \centering
    \includegraphics[width = 0.97\textwidth]{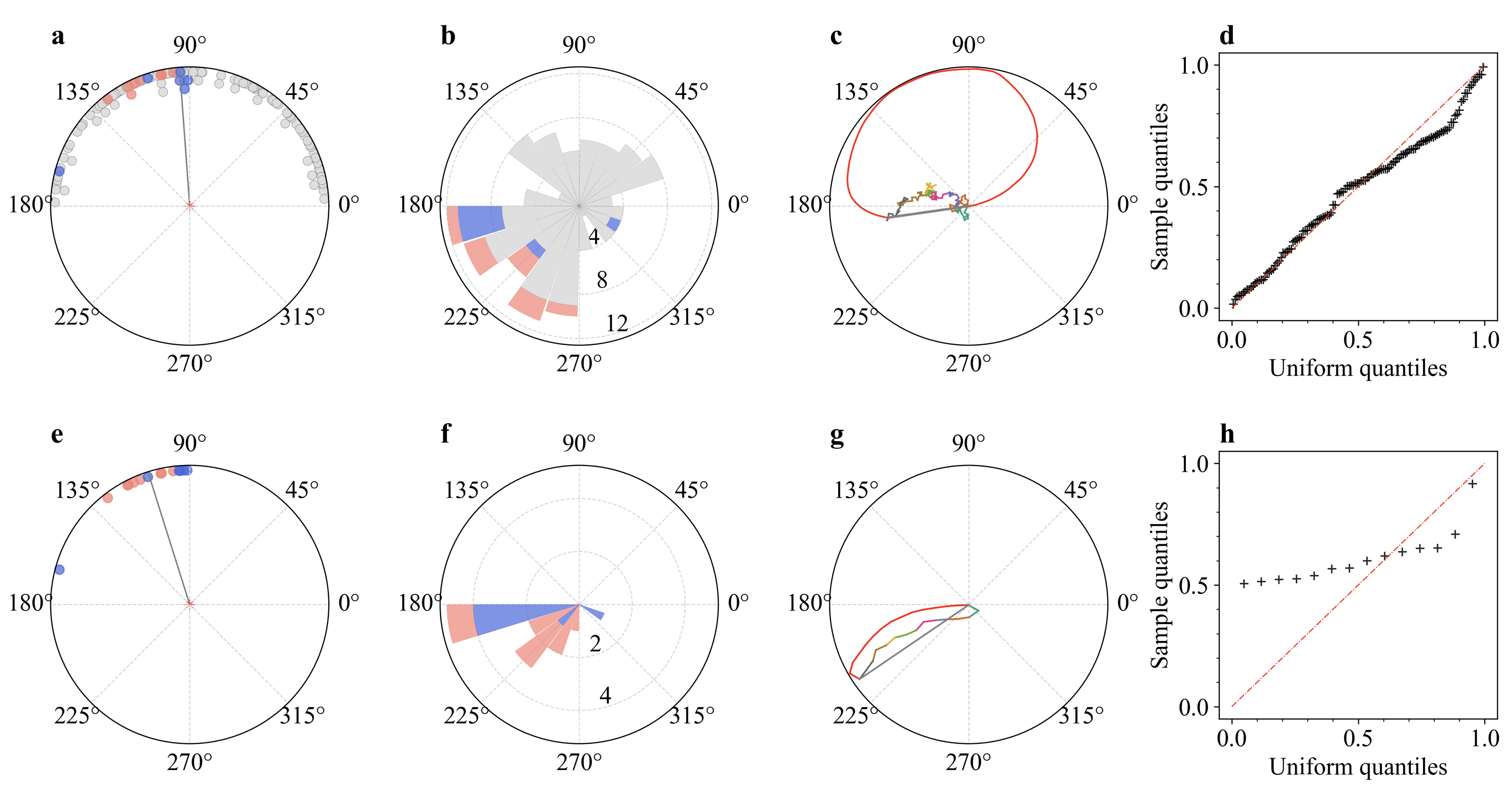} 
    \caption{Finger plots of the GPA (left), rose plots (middle left) and vector plots (middle right) of the doubled 
    GPA, and the quantile–quantile (Q–Q) plots (right) of the measured GPAs for the entire bulge PN sample
    (\textbf{a-d}) but where GPA measurements in the short-period binary sub-sample are color-coded blue for the 
    PNe with confirmed binary and red for the high-\textit{adf}. Just PNe in the binary sub-sample are shown in 
    (\textbf{e}-\textbf{h}) with the same colour code. The mean GPAs are 
    indicated with the black lines in the finger plots. The vector plots in panels (\textbf{c}) \& (\textbf{g}) show 
    the additions of unit vectors in the directions of the doubled GPAs ordered either by the PN Galactic coordinate 
    (in colors) or by the magnitude (in red). Deviation from the red line in the Q-Q 
    plots, which represents a consistency with a uniform distribution, indicates a non-uniformity which is very 
    striking in (\textbf{h}).}
    \label{fig:stats_plot}
\end{figure*}

\section{Discussion and Speculation} 
The bulge PNe currently going through the short-lived PNe phase, regardless of whether in a binary or even triple system 
with other companions (including planets), could have a range of progenitor masses, as suggested by 
\citet{bensby2013chemical, bensby2017chemical}. The progenitors will not have been born at the same time. 
\citet{gesicki2014accelerated} derive an extended star formation history from bulge PNe. For example, a range of 1 
solar mass at stellar birth would mean a difference of hundreds of millions of years for stars born with masses 
between 2 and 3 M$_{\odot}$ \citep[see Fig.~3 in][] {2002AJ....123.1188B} based on evolutionary models of single stars. 
In a PNe hosting a binary or even multiple system, only one of the stars is currently going through the 
PNe phase as it reaches that particular stage of 
evolution. 

By chance, two such PNe in our short-period binary sub-sample have had their CSPN ages previously 
estimated by \citet{2014A&A...566A..48G}, namely Th~3-12 (PNG~356.8+03.3) and H~2-13 (PNG~357.2+02.2), at 7 and 
10 Gigayears respectively (but based on a single-star model).
Hence, if the observed GPAs of their major bipolar lobes do indeed 
reflect the inclinations of their binary orbits today, and if these orbital inclinations originate from system birth, 
then the causal influence acting on star formation has to have endured, effectively 
unchanged, for cosmologically significant time periods.
The difference of about 3~Gigayears just for the two examples above suggests that the alignment process was not caused 
in a single star formation burst but acted for at least several Gyr.

Why do only the bipolars with short-period binaries demonstrate the 
astonishing GPA alignment shown across several kpc of the Galactic bulge? 
A possible explanation that other bipolars are not in (close) binary systems is contrary to currently 
accepted wisdom for how bipolars form e.g. \citet{2022A&A...660L...8O}. 
Our results imply that more than one type of systems forms bipolar nebulae, given only a modest fraction of observed 
bipolars fit the bill. Perhaps the observed alignment effect is also a function of the separation of the binary 
components at system birth? Note that close binary CSPN have isolated binary evolution and the angular momentum 
vector of the system remains constant after formation. 
The only other alternative is a physical process operating now in the 
Galactic bulge but only on bipolar PNe in short period binaries, that aligns their bipolar ejecta over the 
short timescales that these PNe are now visible (only a few tens of thousands of years).

PNe axes of symmetry are thought to trace the rotation of their CSPN which may themselves align with 
a very strong interstellar magnetic field at their time of formation from the collapsing disks of rotating 
molecular gas where the orbital inclinations, if also part of a binary or multiple system, are established 
\citep{rees2013alignment}. An alternative is they may possibly be directly tilted later by the extant 
magnetic field itself \citep{falceta2014alignment}. Binary star systems may have formed long ago
in the presence of a strong Galactic magnetic field where their orbits align with this field due to 
their higher angular momenta compared with single-star systems \citep{rees2013alignment}. Indeed, it 
is reported by \citet{2007MNRAS.377...77P} that magnetic fields have important effects on molecular cloud core 
fragments and on their subsequent condensation and formation of binary systems. They report the scale of binary 
separation at birth as an important factor in the level of binary orbit to magnetic field alignment eventually 
achieved with the effect more pronounced at shorter separations.

Symmetry axes of PNe ejecta from one of the stars within binary systems take the axes of their
angular momenta. The magnetic field in the Galactic bulge was proposed to counter the contraction of 
any star-forming cloud perpendicular to the direction of the Galactic plane during the early phases of 
bulge star formation \citep[e.g.][]{2007MNRAS.377...77P, 2018MNRAS.473.2124G}. However, we observe the 
preferential major axis orientation 
only in objects with confirmed or inferred 
short-period binaries that have also likely undergone a CEE phase. These objects may have been in 
tighter orbits to begin with before then spiralling in for the CEE phase and may have higher 
angular momentum. PNe that avoid a CEE would have much wider orbits at birth. 

According to \cite{falceta2014alignment} PNe symmetry axes modification can occur if the local 
magnetic field is $>100$ $\mathrm{\mu G}$.  From optical polarization measures, radio
synchrotron data and Zeeman splitting seen in object spectra, the average total 
magnetic field strength in the bulge today is  $\sim$20-40 $\mu$G \citep{larosa2005evidence}. This is likely 
insufficient to modify binary orbital inclinations over the short timescales implied before the bulge PNe we
see today are formed. It is more plausible the observed effect was 
``frozen in" in the past. The simulations of \citet{falceta2014alignment} 
show further orbit modification can occur at t~$<$ 10$^{4}$ years after PN envelope 
ejection, compared with the rapid CEE timescaled expected \citep[a few years or even less, 
][]{ivanova2013common}. This might not happen to post-CEE PNe that form our short period binary 
sample. Only post-CEE PNe may maintain their ancient orientations resulting from their orbital angular 
momenta  aligning with the initial (presumably historically stronger) magnetic 
field at their formation time along the Galactic plane. This assumes they originally had tighter orbits 
following the arguments of \citet{2007MNRAS.377...77P}. 

PNe formed from wider binaries 
could have their symmetry axes change over time more randomly due to the ambient field where 
directions may evolve due to a Galactic wind or bulk motions of the ISM. This provides a possible 
explanation for the observed effect only being seen in post-CE binary PNe. If this is the process then it must remain ordered, stay potent and have endured for billions of years. The only alternative is a pervasive force (magnetic?) acting currently only on the bipolar lobe ejection mechanism in PNe hosting short period binaries that is sufficiently strong to align GPA's over the entire Galactic bulge in a 
few thousand years.
\begin{acknowledgments}
\section*{Acknowledgements}
We are grateful to the referee and statistics  editor whose comments and suggestions have significantly improved the paper. S.T. thanks HKU and Q.A.P. for provision of an MPhil scholarship and subsequent Research
Assistant position, Q.A.P. thanks the Hong Kong Research Grants Council for GRF research support under grants 17326116 and 17300417. 
A.R. thanks HKU for the provision of postdoctoral fellowship under Q.A.P. A.A.Z. thanks the Hung Hing Ying Foundation for a visiting HKU professorship and acknowledges UK STFC funding under grant ST/T000414/1. 
This work made use of the University of Hong Kong/Australian Astronomical Observatory/Strasbourg Observatory
H-alpha Planetary Nebula (HASH PN) database, hosted by the Laboratory for Space Research at the University of Hong Kong. 
We acknowledge use of ESO observations under program IDs 095.D-0270(A), 097.D-0024(A), 099.D-0163(A), and 0101.D-0192(A) for PI Rees. We acknowledge use of data from the Wide Field and Planetary 
Camera 2 Instrument (WFPC2) of the \emph{HST} through HST-SNAP-8345 (PI: Sahai), HST-SNAP-9356 (PI: Zijlstra) and HST-GO-11185 (PI: 
Rubin). Some data presented in this paper were obtained from the Mikulski Archive for Space Telescopes (MAST) at the Space Telescope 
Science Institute. The specific observations analyzed can be accessed via\dataset[doi:0.17909/pej3-ce46]
{https://doi.org/10.17909/pej3-ce46}.
\end{acknowledgments}
\facilities{\emph{HST} (WFPC2), ESO VLT (FORS2)}
\software{APLpy \citep{robitaille2012aplpy, robitaille2019aplpy}, Astropy \citep{astropy:2018}, circbayes \citep{circbayes}, CircStats 
\citep{berens2009circstat}, emcee \citep{foreman2013emcee,foreman2019emcee}, Matplotlib \citep{hunter2007matplotlib}, numpy \citep{van2011numpy}, Python \citep{van1995python}, R \citep{R_language}, scipy \citep{2020SciPy-NMeth}, sphunif \citep{sphunif}.}


\bibliography{main}
\section*{Appendix A: PN Images and GPAs}
\label{append:img_gpa}
\renewcommand{\thefigure}{A1}
\begin{figure}[!h]
    \centering
    \caption{Grey-scale images on an arbitrary scale of the PNe sample. Each plot has North at the top and East at the left. 
    Arcsecond lengths are indicated by lines at the upper left. Panel \textbf{(a)} presents that with both a confirmed 
    short-period binary central star and a measurable EPA/GPA, together with M~1-34, suspected of hosting a short period binary. 
    Our best choice for the axis of long symmetry is indicated by a blue dashed except for M~1-34 which is shown with a green 
    dashed line with measured EPAs shown in the upper right in blue. The PN normal name and the converted GPA are shown at the 
    bottom left of each PN image. Panel \textbf{(b)} shows that of the eight PNe with both a high \textit{adf} and a 
    measurable EPA, following the same notation adopted for \textbf{(a)}. Our best choice for the axes of long symmetry 
    are indicated by red dashed lines.}
    \label{fig:PA_binary}
    \begin{tabular}{cc}
        \multicolumn{2}{c}{\textbf{(a)}}\vspace{0.15cm}\\
    \end{tabular}\\
    \includegraphics[width = 0.233\textwidth]{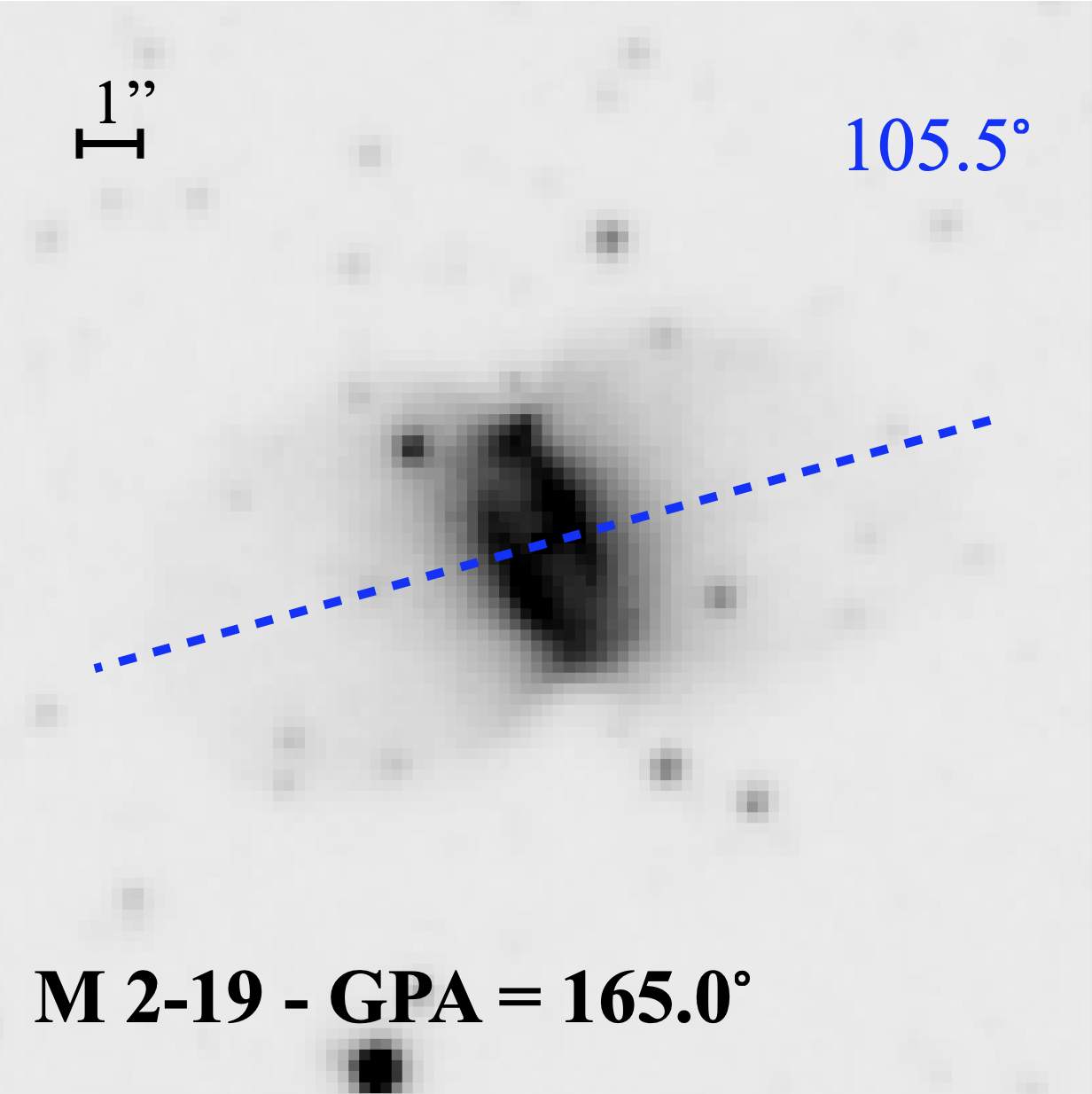}
    \includegraphics[width = 0.233\textwidth]{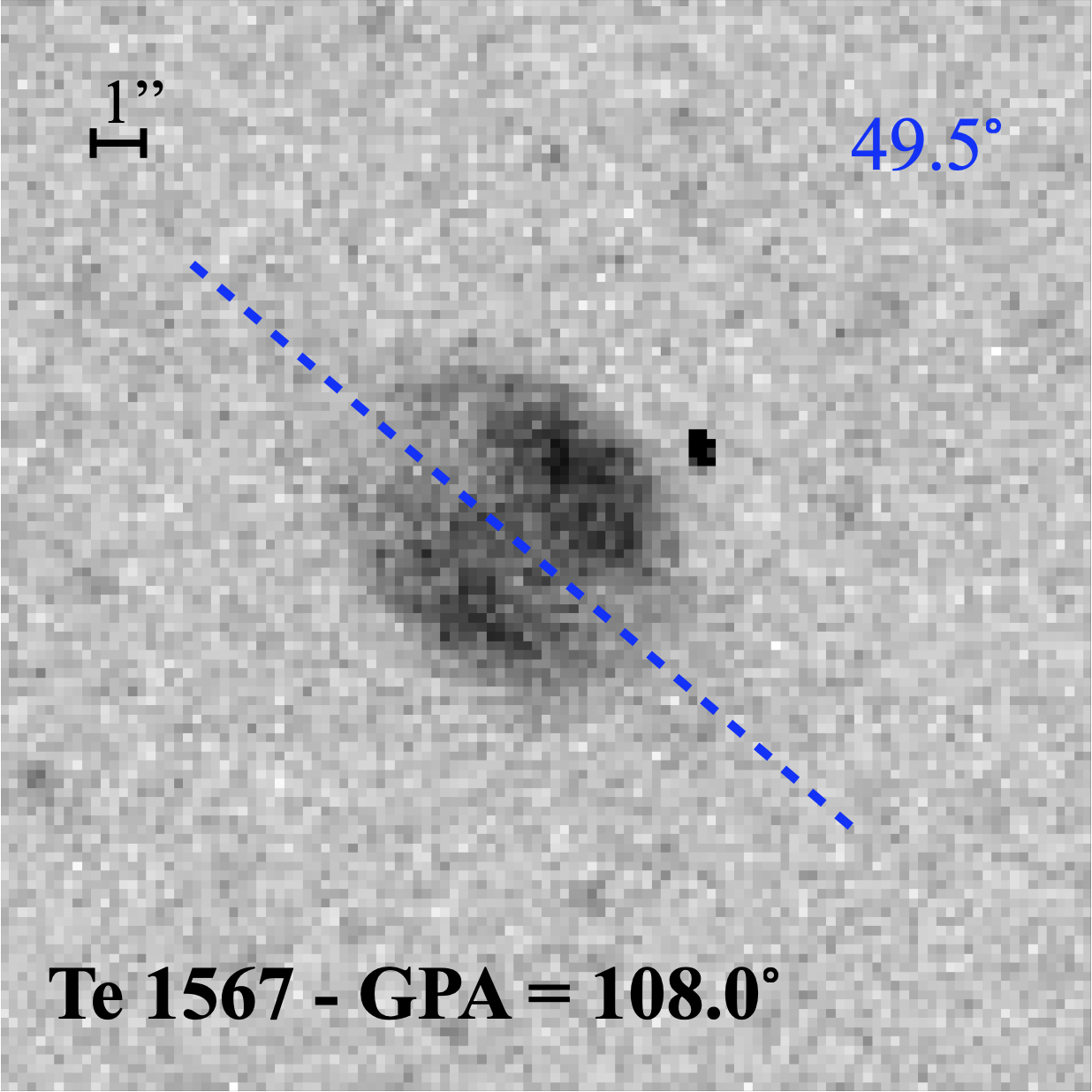}
    \includegraphics[width = 0.233\textwidth]{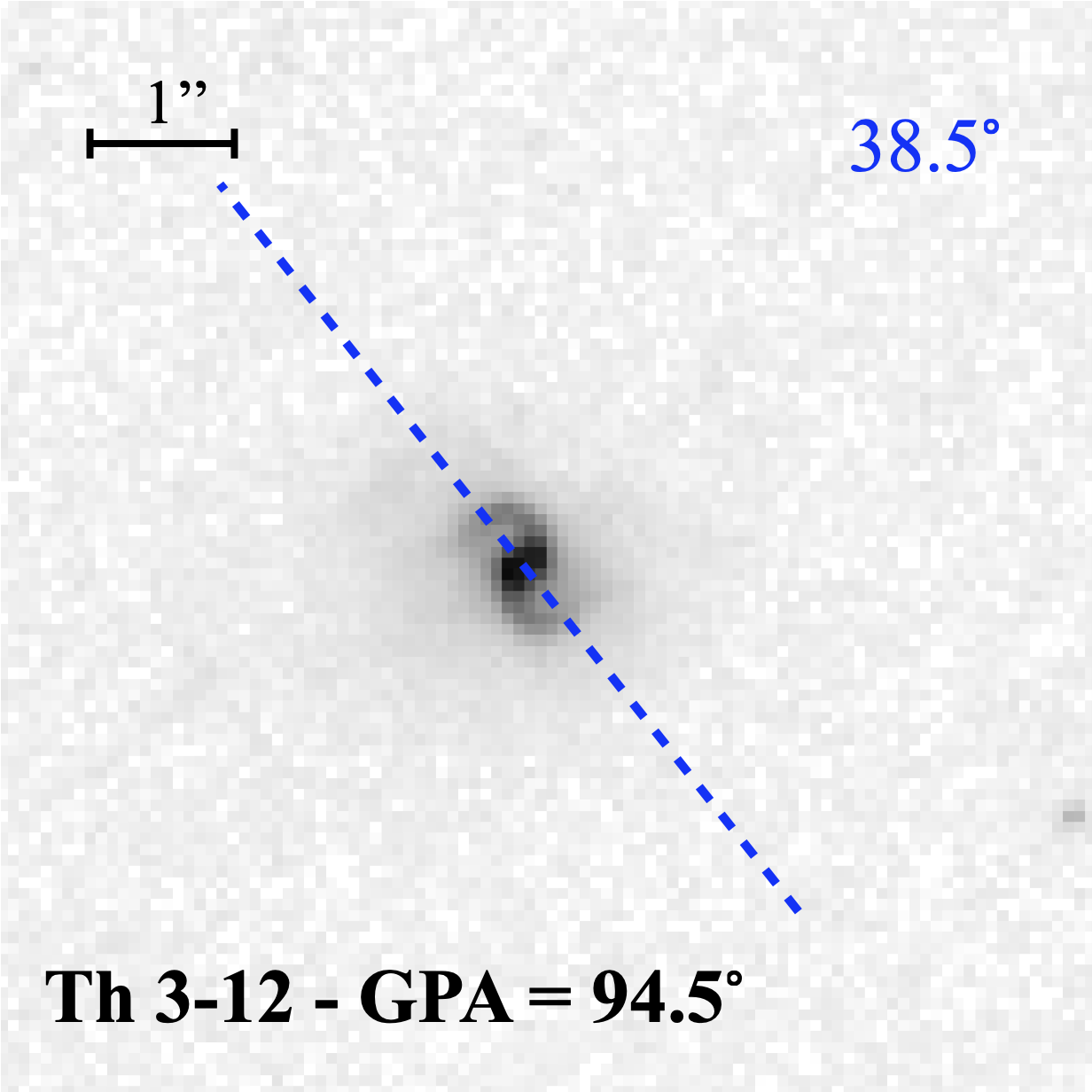}
    \includegraphics[width = 0.233\textwidth]{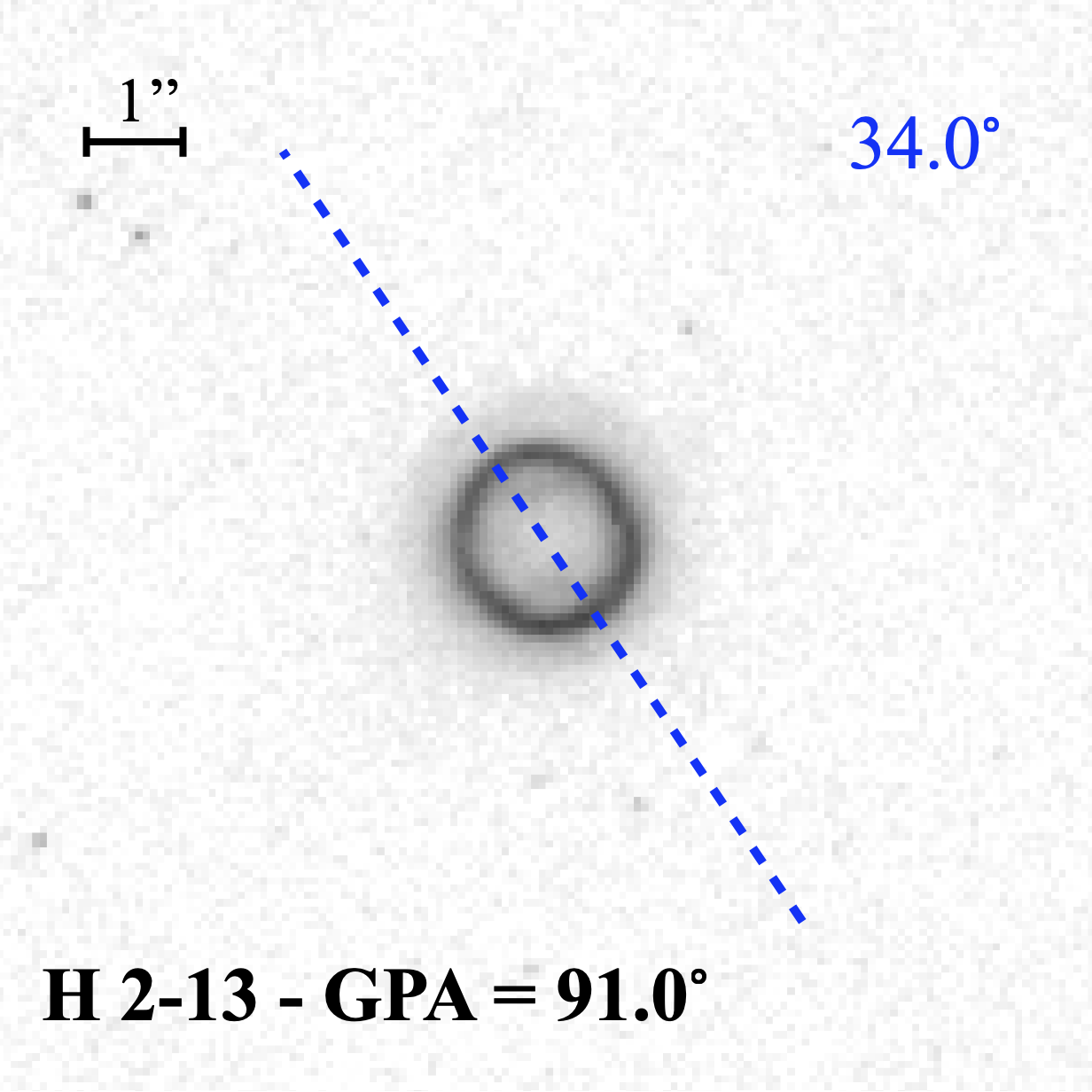}
    \includegraphics[width = 0.233\textwidth]{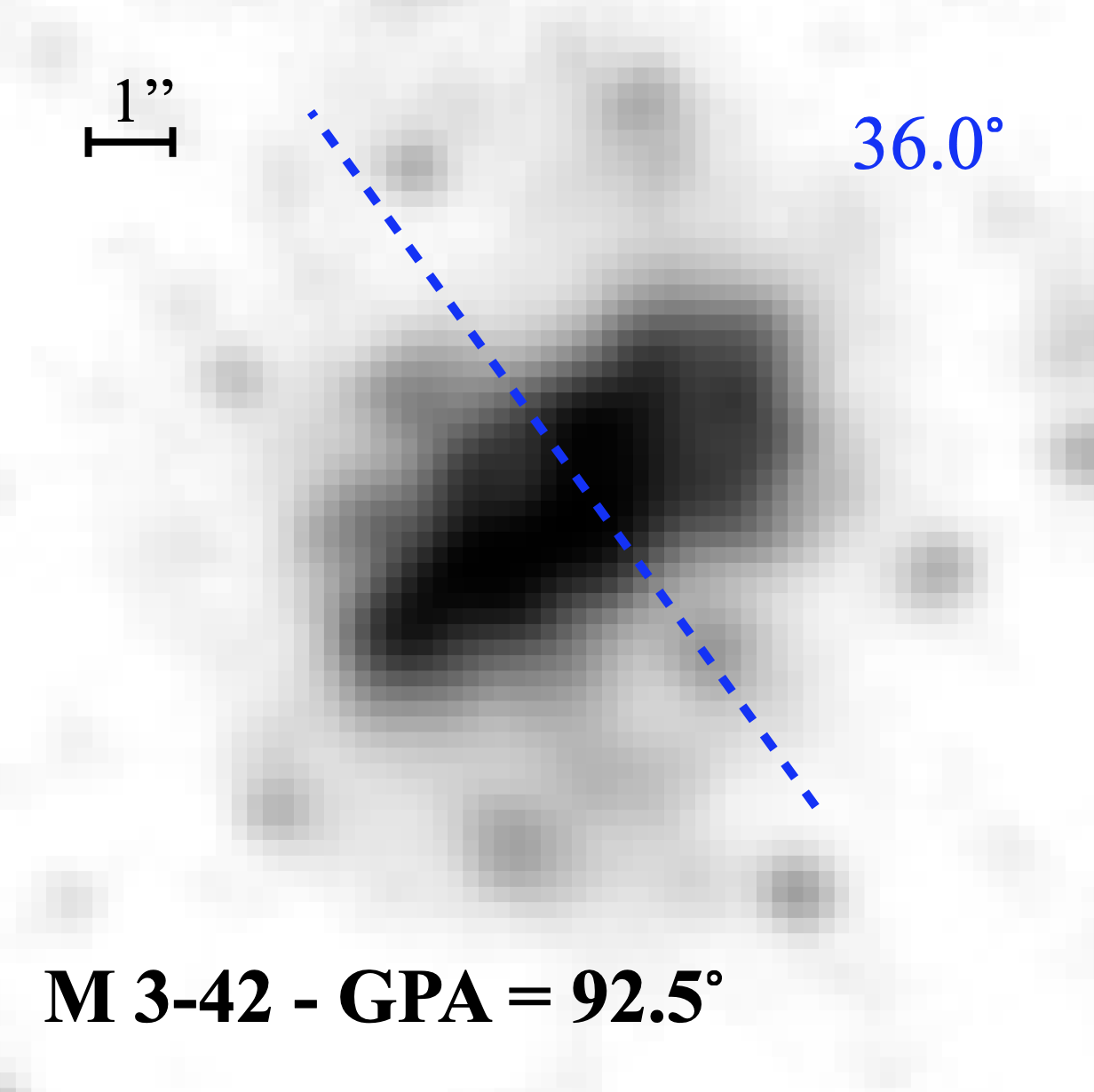}
    \includegraphics[width = 0.233\textwidth]{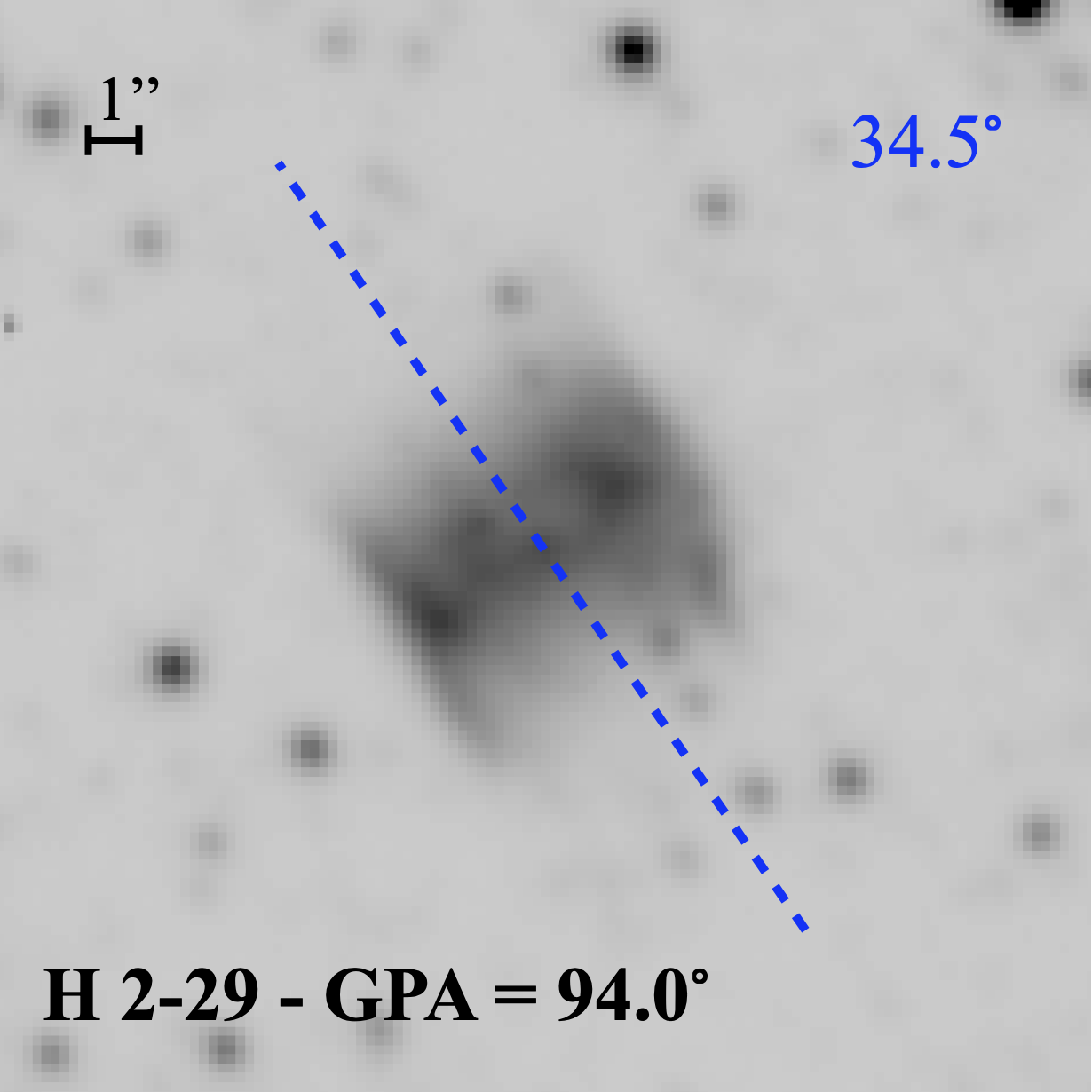}
    \includegraphics[width = 0.233\textwidth]{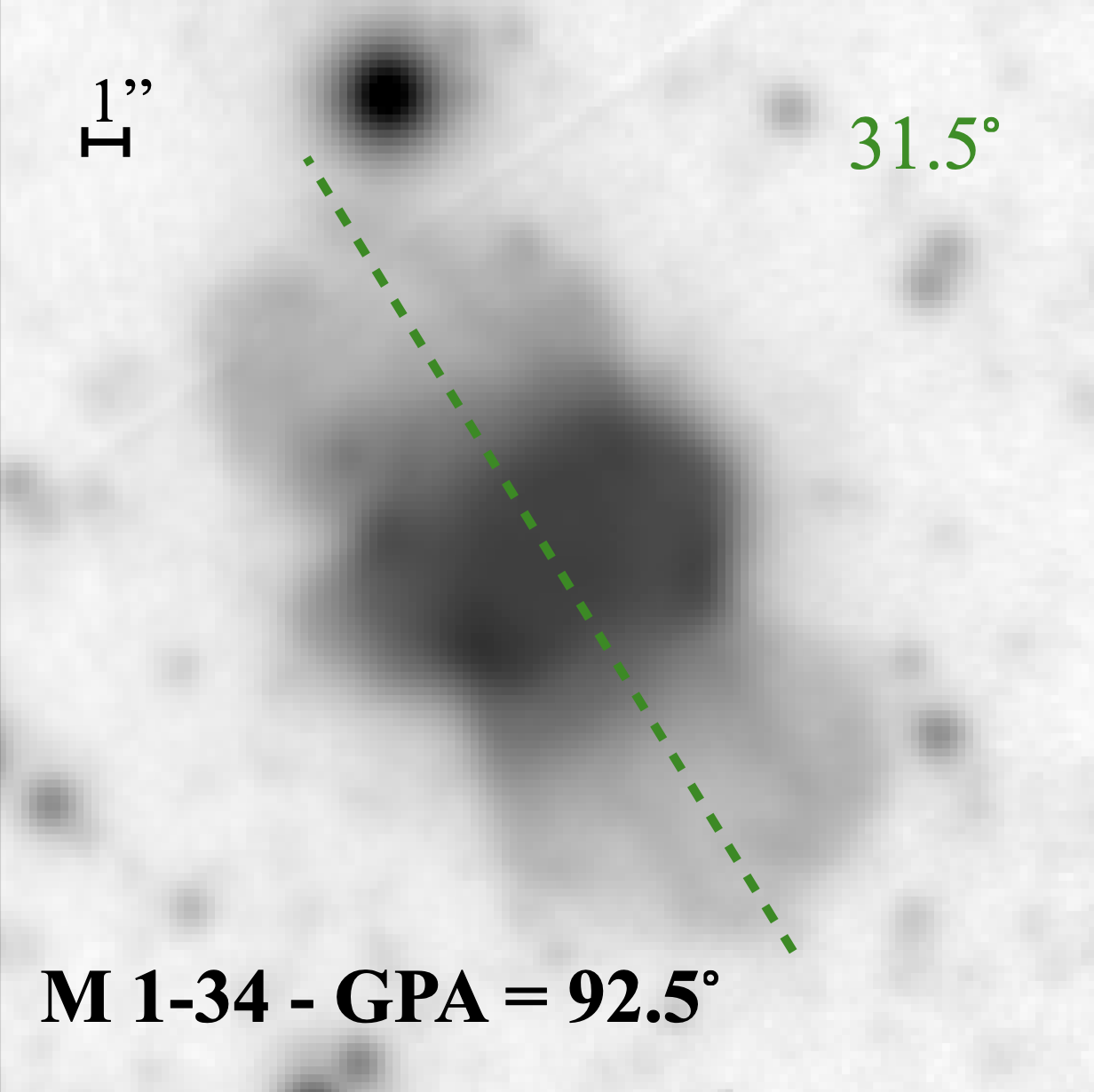}
    \newline
    \begin{tabular}{cc}
    \\
    \multicolumn{2}{c}{\textbf{(b)}}\vspace{0.15cm}\\
    \end{tabular}\\
    \includegraphics[width = 0.233\textwidth]{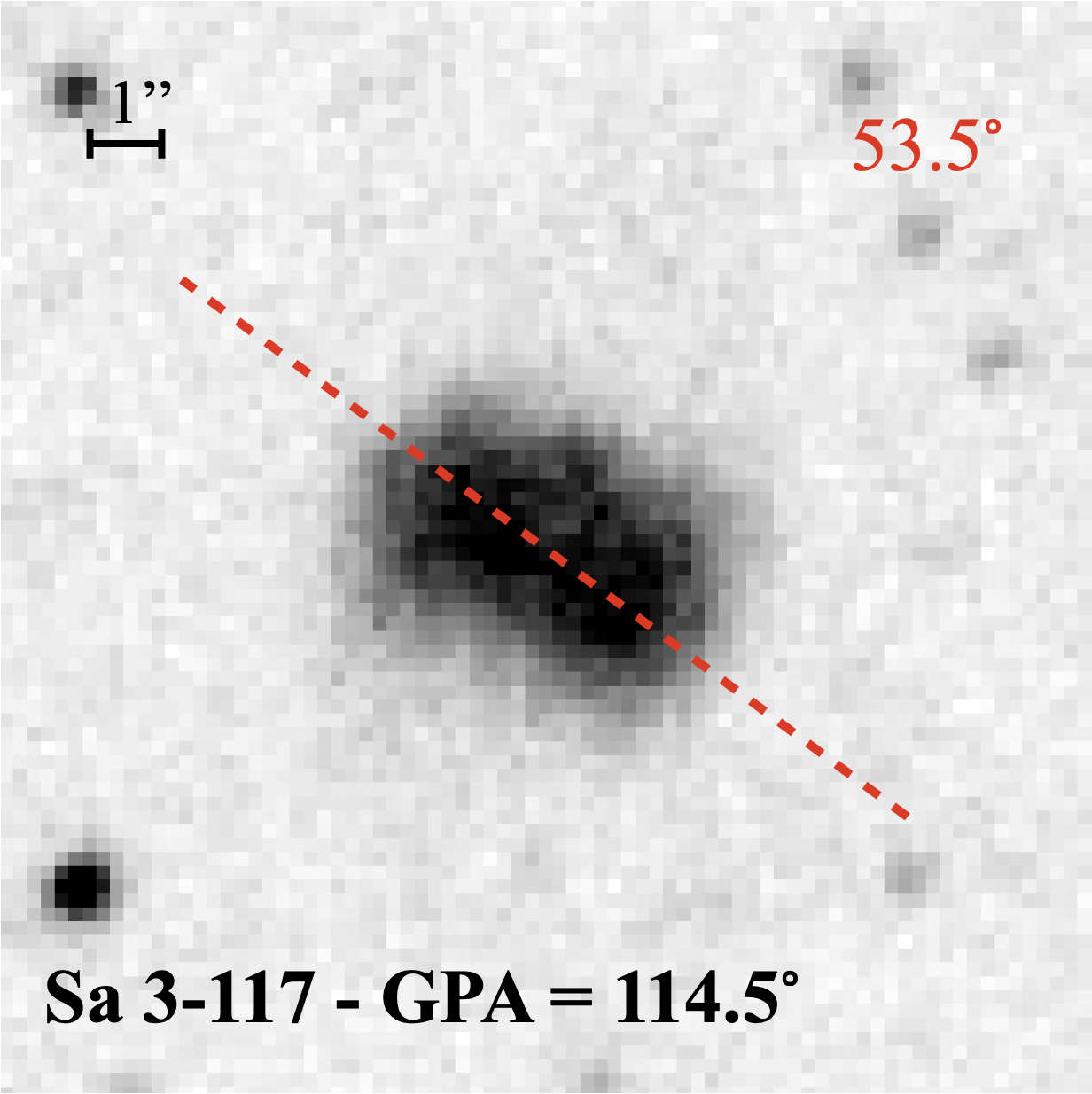}
    \includegraphics[width = 0.233\textwidth]{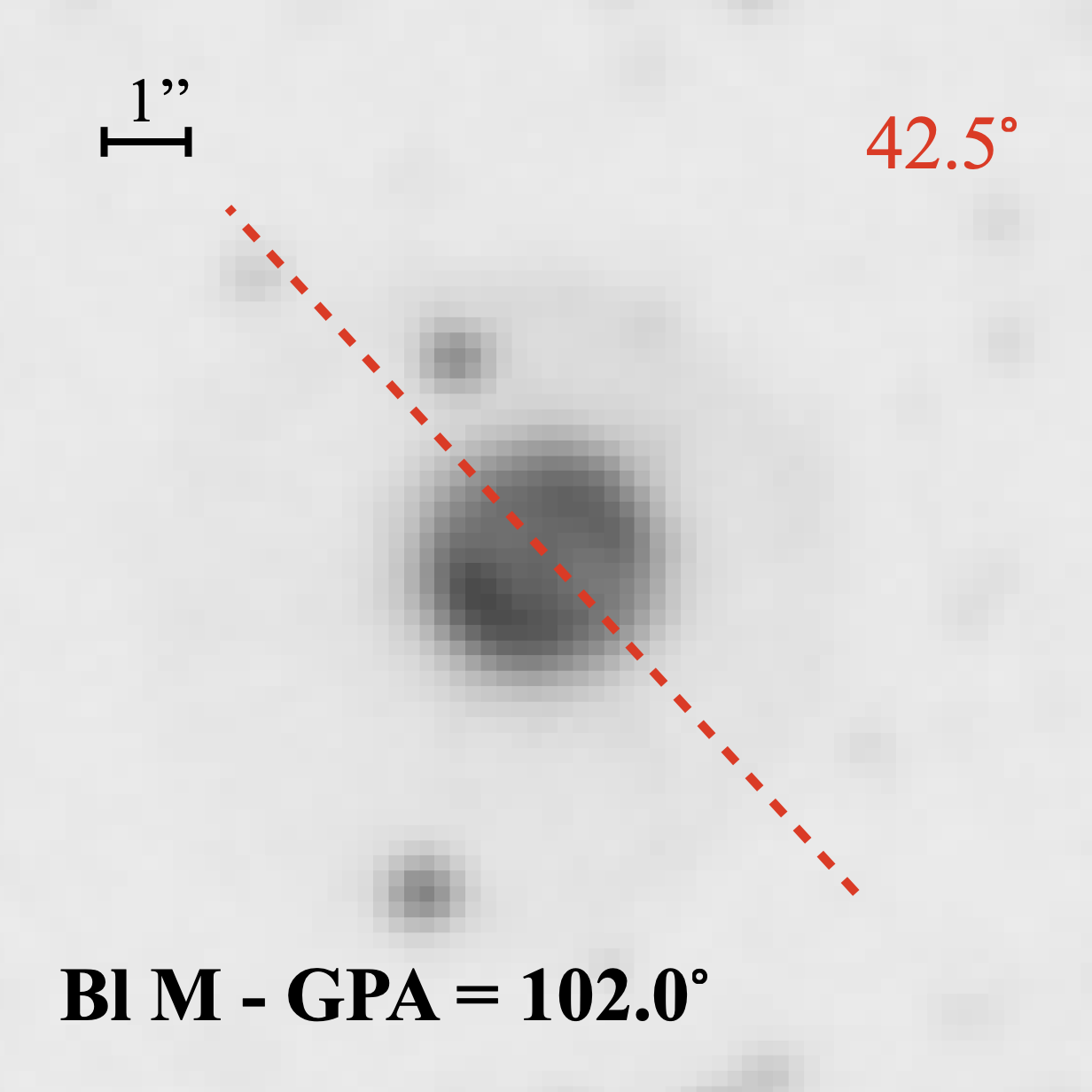}
    \includegraphics[width = 0.233\textwidth]{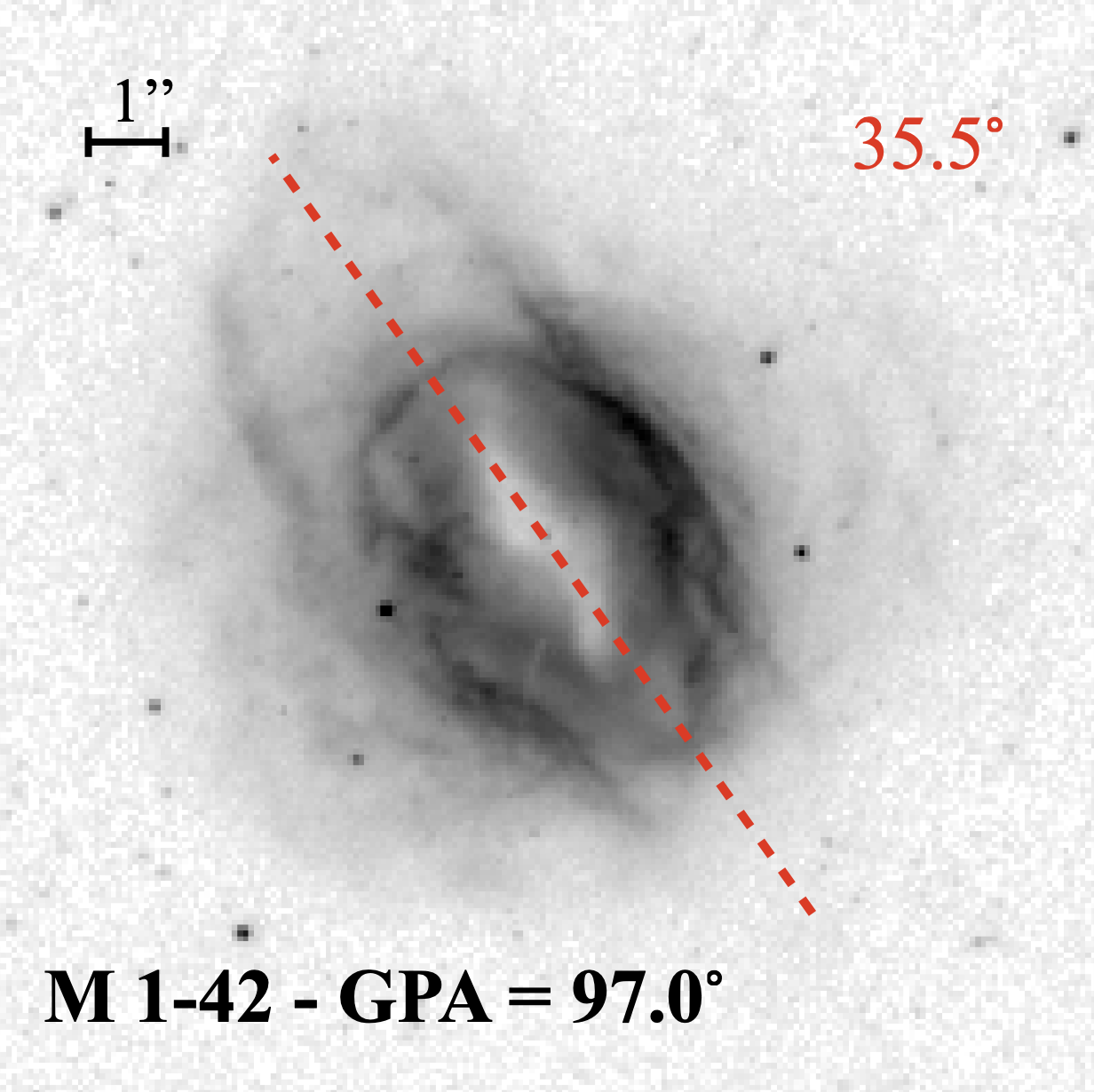}
    \includegraphics[width = 0.233\textwidth]{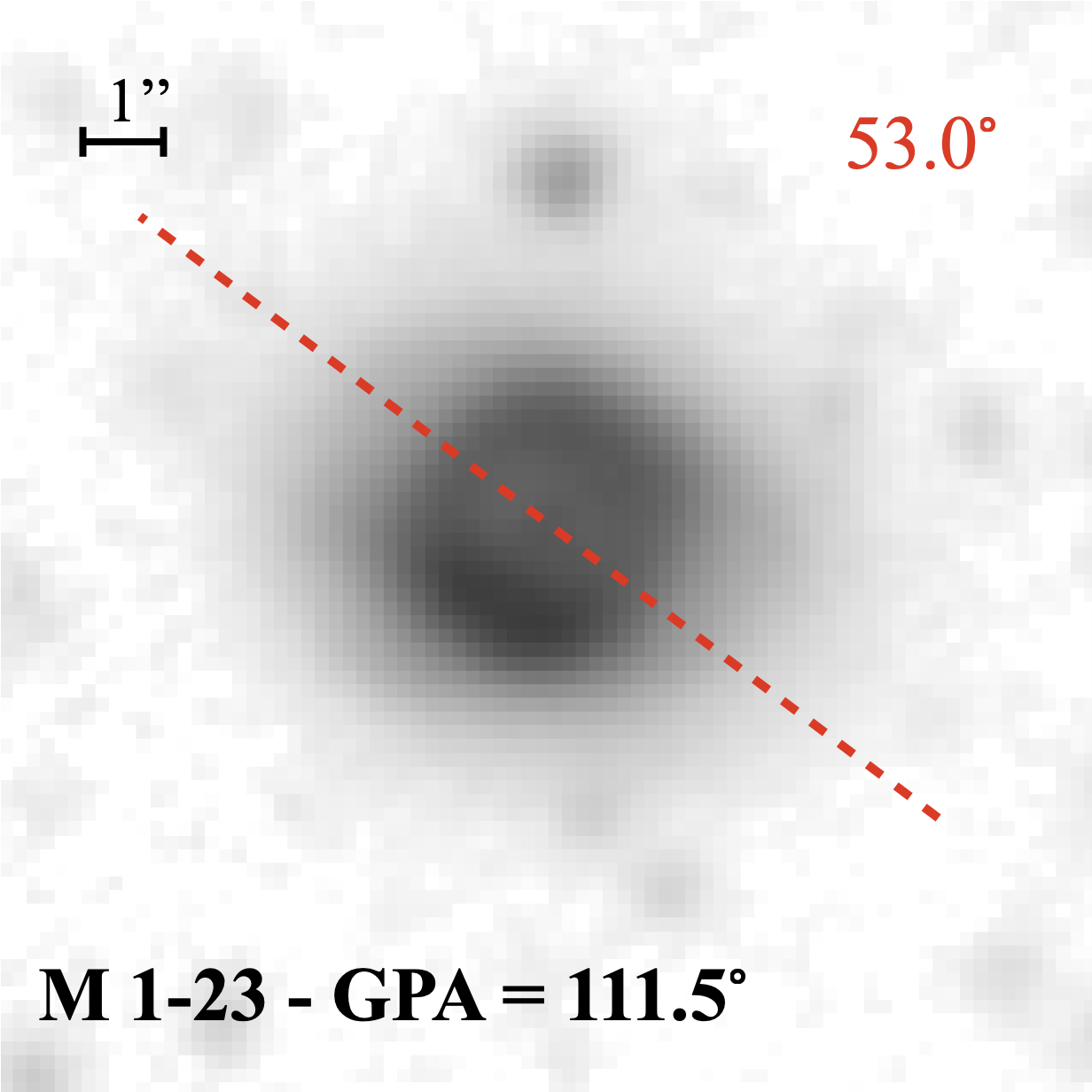}
    \includegraphics[width = 0.233\textwidth]{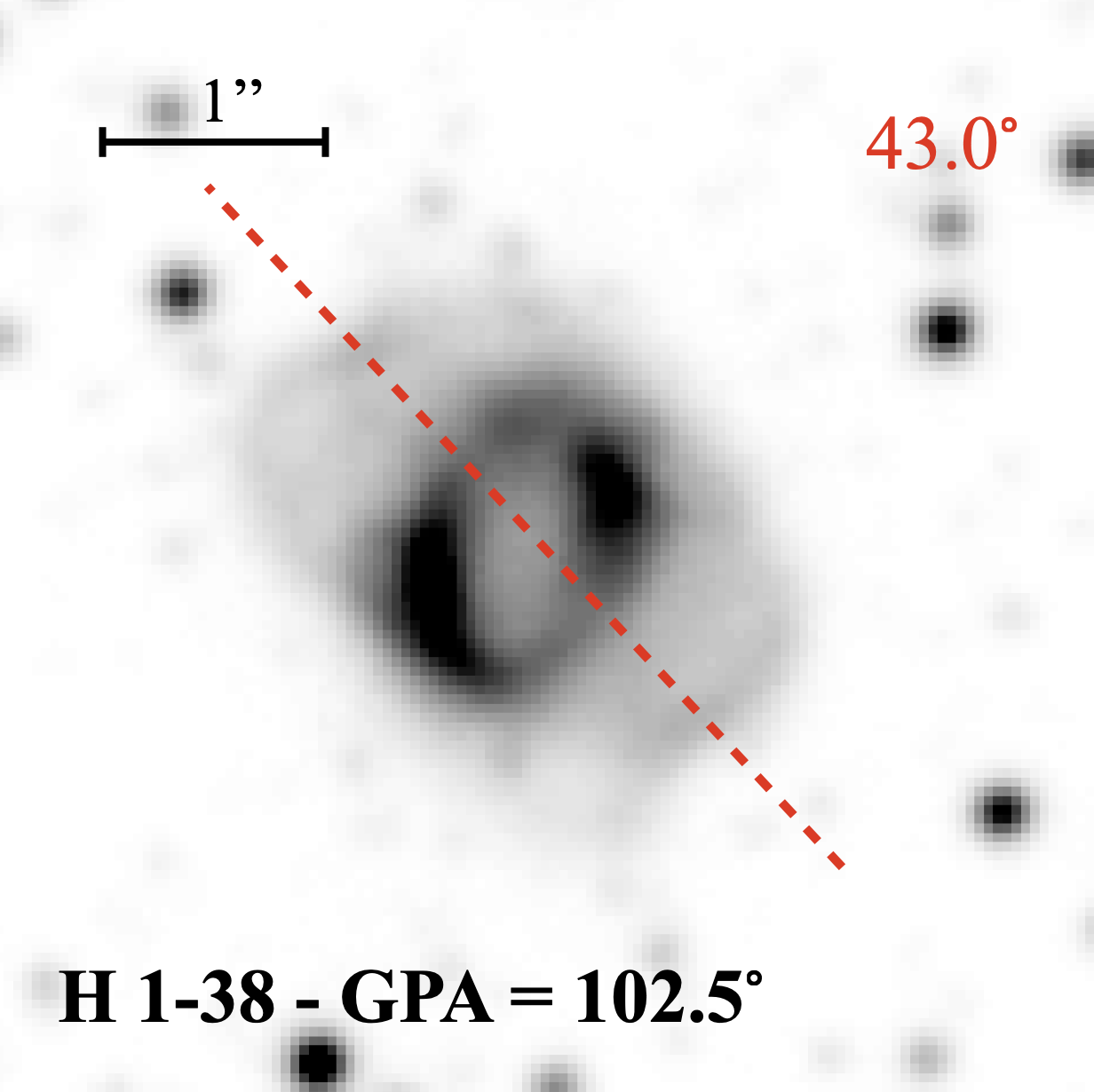}
    \includegraphics[width = 0.233\textwidth]{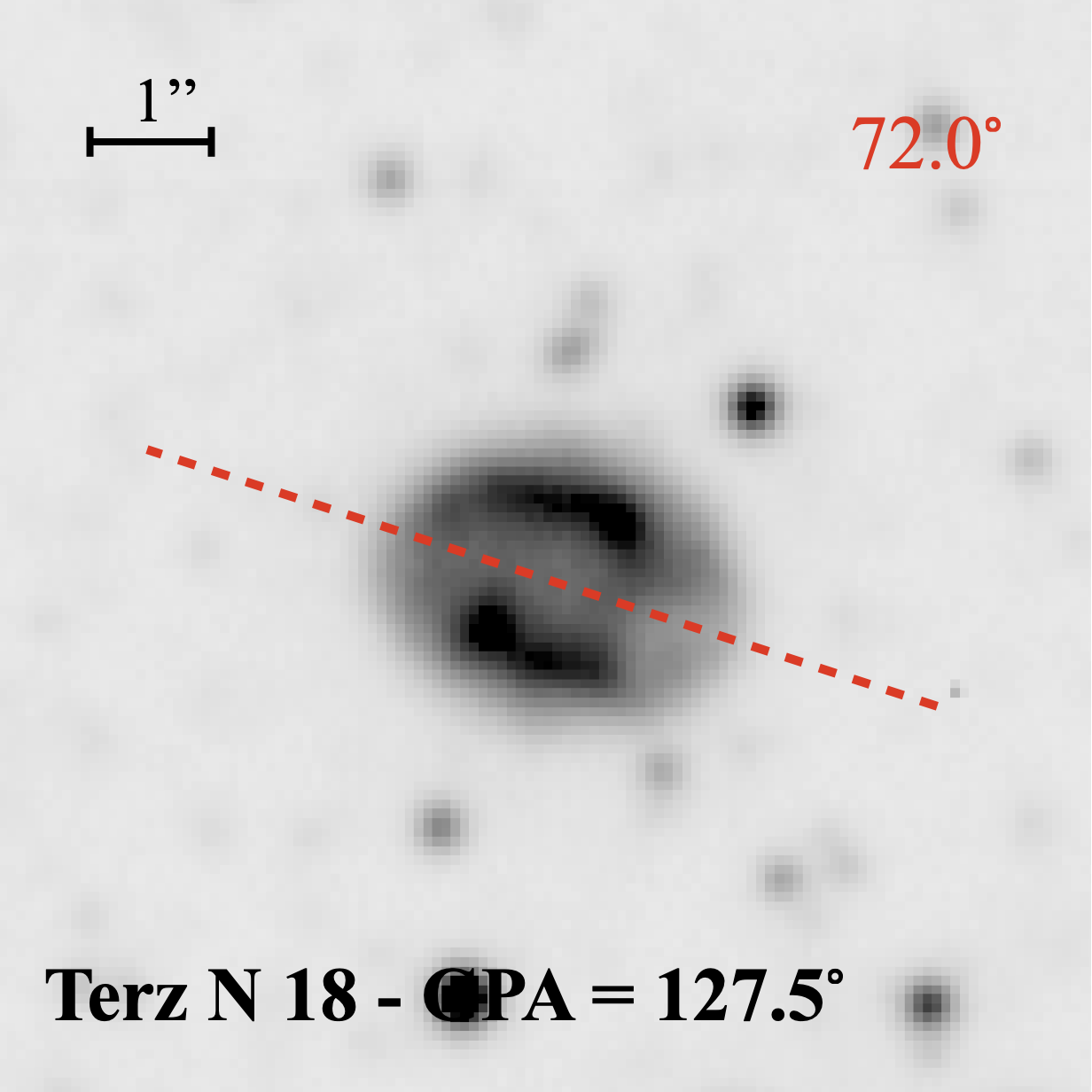}
    \includegraphics[width = 0.233\textwidth]{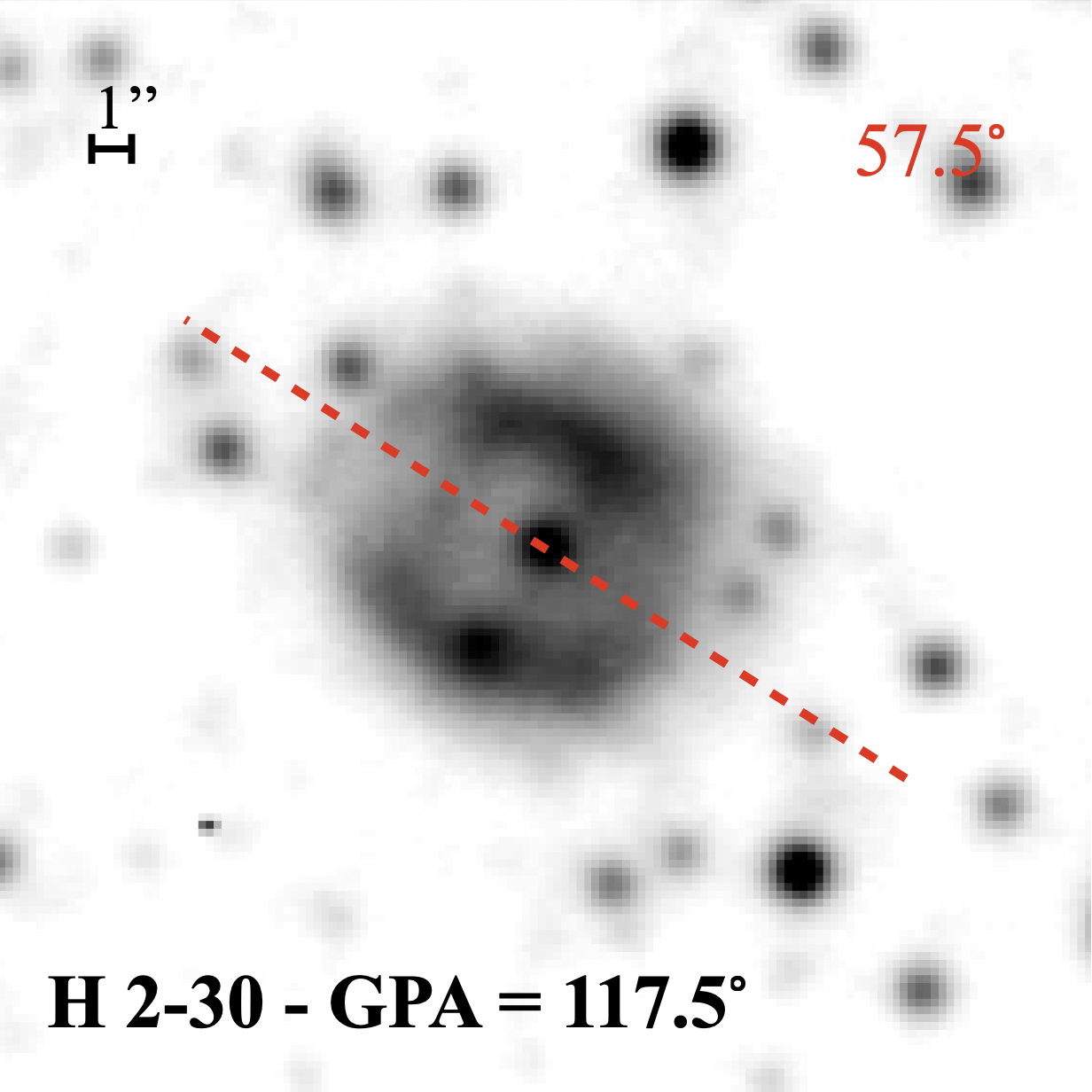}
    \includegraphics[width = 0.233\textwidth]{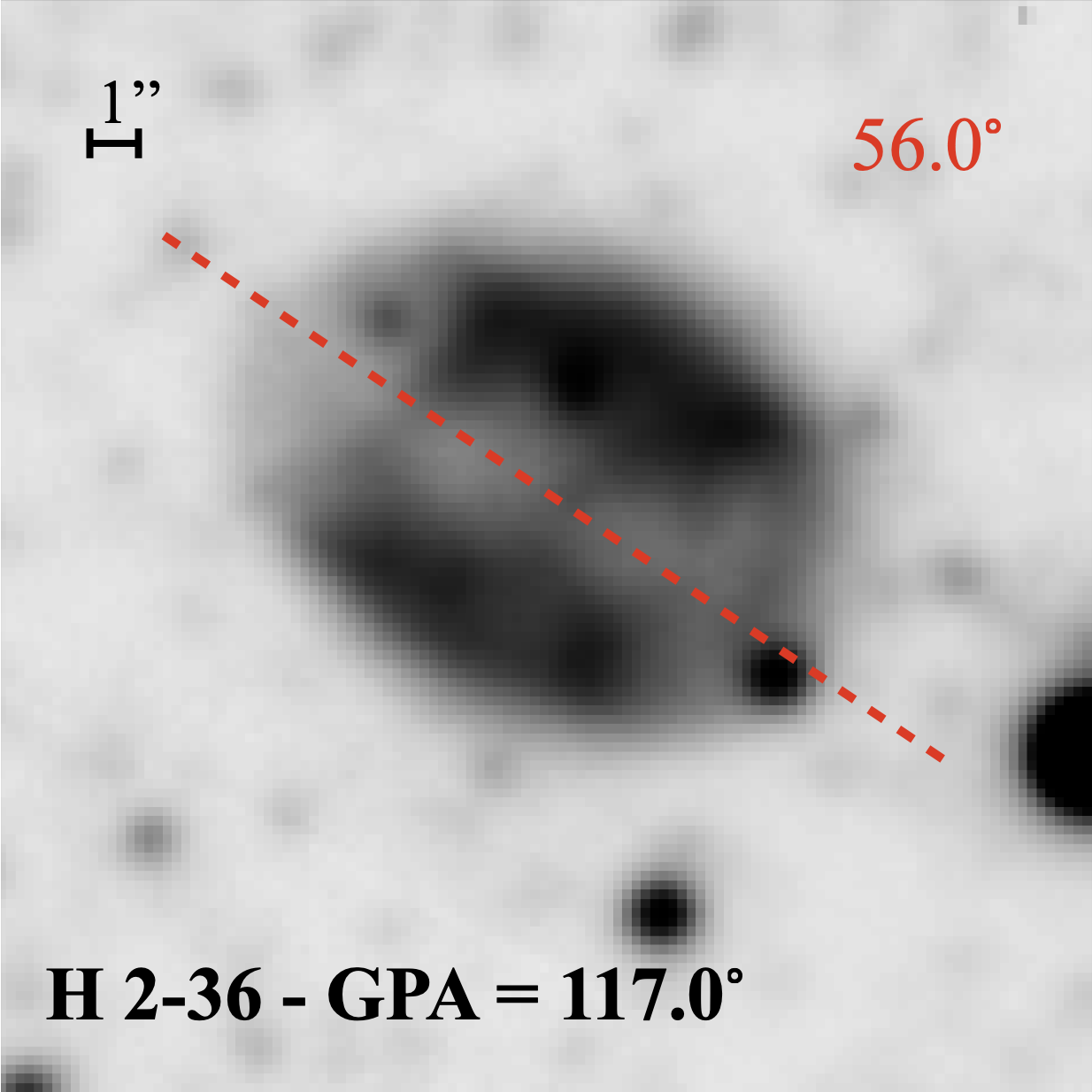}
\end{figure}
\section*{Appendix B: Posterior distributions of von Mises parameters obtained by applying \texttt{emcee} on our short-period binary PN sample.}
\label{append:stats}
\renewcommand{\thefigure}{B1}
\begin{figure}[!h]
\centering
\caption{The so called ``corner plot'' showing the results of MCMC parameter estimation of the von Mises model applied 
to our short-period binary PN sample. The histograms show the marginalised posterior densities for the concentration 
parameter, $\kappa$ (top left panel), and the mean GPA direction, $\theta_{\mathrm{\mu}}$ (bottom right panel), 
 both measured in terms of radians. 
The best-fit parameters $\theta_{\mathrm{\mu}}=1.90\pm0.09$ rad ($=109.0\pm4.9^{\circ}$) and $\kappa=10.39^{+4.21}_{-3.29}$, 
equivalent to $\sigma=15.0^{+6.5}_{-2.8}$$^{\circ}$, are indicated by vertical red lines and displayed on the top of the histograms. The 
distribution of $\kappa$ is skewed towards higher concentrations, indicating an enhanced probability of low GPA dispersion. The dashed lines 
indicate the 16th and 84th percentiles of the posterior samples, providing an estimate of the uncertainties. The MCMC chain used 500 samplers 
and ran for 10,000 steps with a 6500 burn-in period. The figure was created with \texttt{corner.py} \citep{foreman2016corner}.}
\includegraphics[width = 0.50\textwidth]{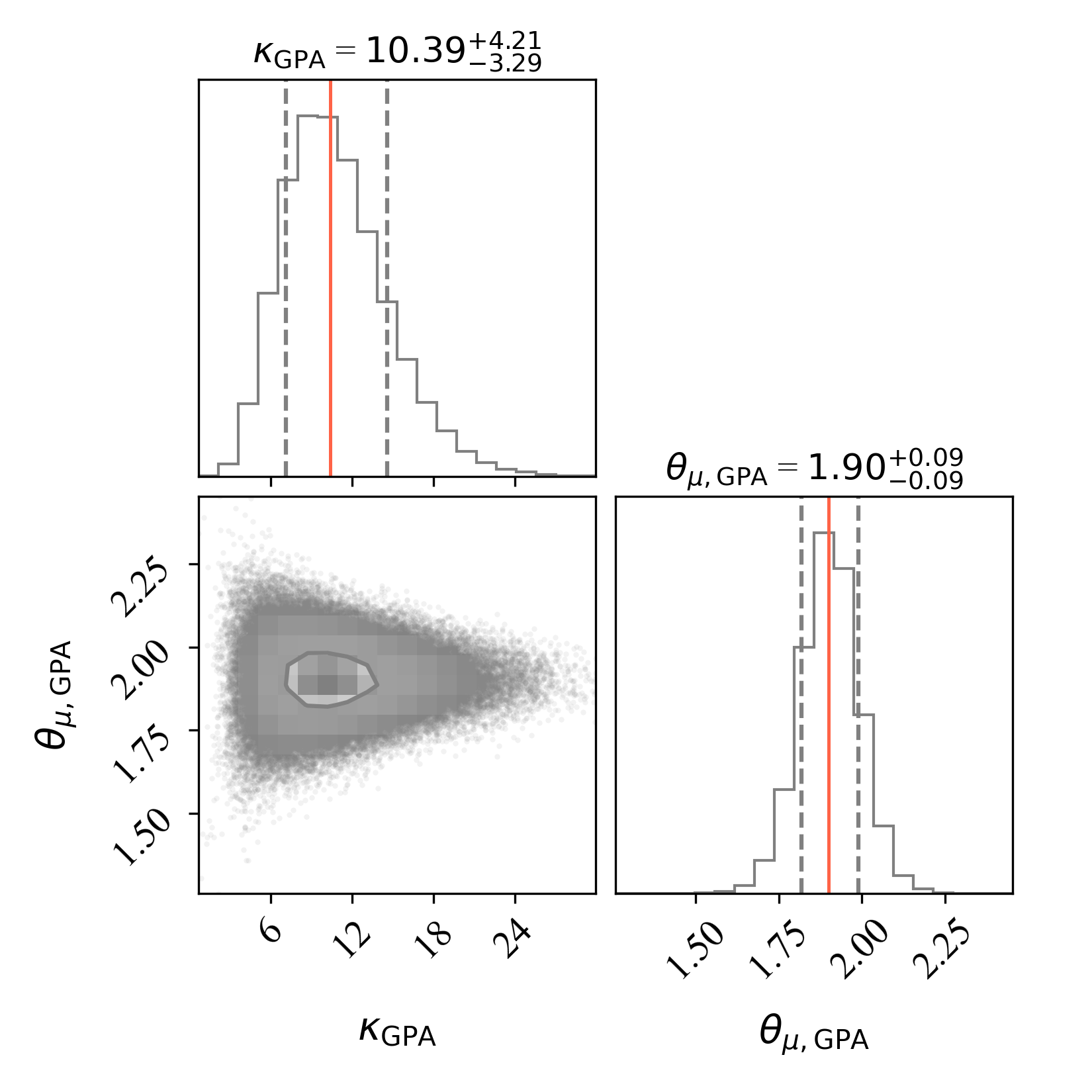}
\label{fig:corner_plot}
\end{figure}
\vspace{-0.5cm}
\renewcommand{\thefigure}{B2}
\begin{figure}[!h]
\centering
\caption{Comparison between probability densities and cumulative distribution functions of the resulting von Mises 
distributions from the MCMC analysis and the observed data from our short-period binary PN sample. 
The left panel shows the observed data as histograms with a bin size of 4$^{\circ}$, overlaid by the probability 
density function of a von Mises distribution with parameters from each of 50 random draws from the fitted posterior, 
represented by grey lines. The best-fit von Mises distribution is highlighted in red. The right panel displays 
the corresponding cumulative distribution functions of the von Mises distributions from the MCMC analysis using 
the same notation as the left panel, with the empirical cumulative distribution function of the observed data 
indicated by the black line. Overall, the probability densities and cumulative distribution functions of the 
sampled parameters provide a good visual match to the overall distribution of the observed data. However, we 
noticed that there are excessive objects at around 90$^{\circ}$ and 160$^{\circ}$ that deviate 
from the best-fit unimodal distribution.}
\includegraphics[width = 0.92\textwidth]{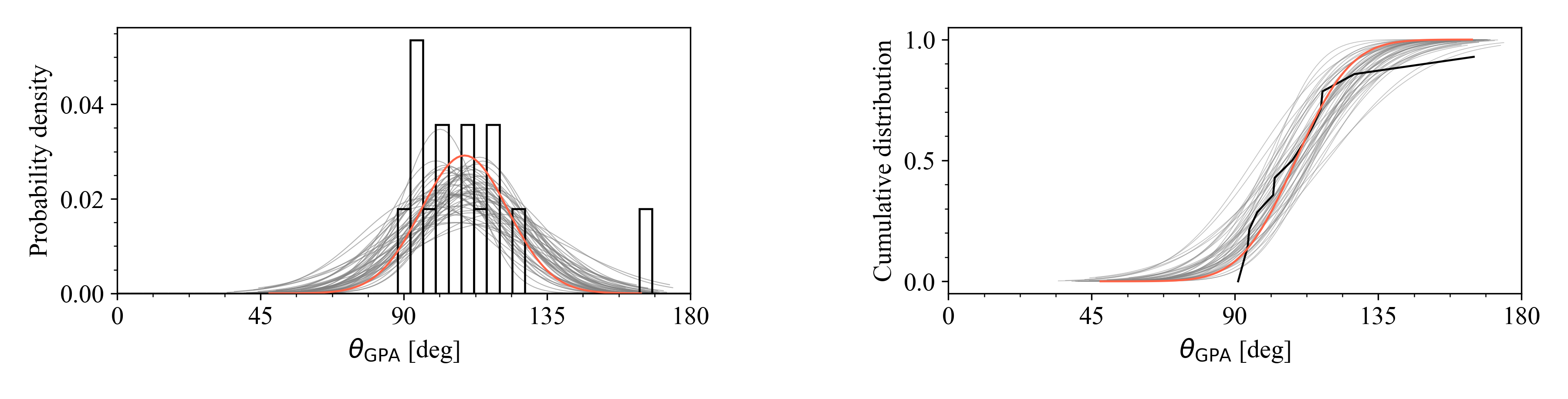}
\label{fig:mcmc_fits}
\end{figure}
\end{document}